\documentclass{IEEEtran}
\usepackage{algorithm}
\usepackage{algpseudocode}
\usepackage[utf8]{inputenc}

\usepackage{amsmath,amssymb,epsfig,psfrag,cite,subfigure}
\include{macros}
\usepackage{graphicx}
\usepackage{epstopdf}

\usepackage{acronym}

\usepackage{bm}

\usepackage{geometry}
\usepackage{tcolorbox}

\usepackage{accents}

\allowdisplaybreaks

\usepackage{soul}

\usepackage{amsthm}

\usepackage{nicefrac}

\usepackage{lipsum,multicol}

\usepackage{algorithm}
\usepackage{algpseudocode}

\algdef{SE}[SUBALG]{Indent}{EndIndent}{}{\algorithmicend\ }%
\algtext*{Indent}
\algtext*{EndIndent}

\usepackage{enumitem}
\usepackage{mwe}    
\usepackage{epsfig,psfrag}
\usepackage{subfigure}
\usepackage{color}
\usepackage{url}
\usepackage{mathtools,xparse}

\usepackage{gensymb}

\usepackage[multiple]{footmisc}

\usepackage{xr}

\makeatletter
\newcommand*{\addFileDependency}[1]{
  \typeout{(#1)}
  \@addtofilelist{#1}
  \IfFileExists{#1}{}{\typeout{No file #1.}}
}
\makeatother

\newcommand*{\myexternaldocument}[1]{%
    \externaldocument{#1}%
    \addFileDependency{#1.tex}%
    \addFileDependency{#1.aux}%
}

\myexternaldocument{supp}

\usepackage{xcolor}

\usepackage[T1]{fontenc}
\usepackage[utf8]{inputenc}
\usepackage{pgfplots}
\usepackage{grffile}
\pgfplotsset{compat=newest}
\usetikzlibrary{plotmarks}
\usepgfplotslibrary{patchplots}
\usepackage{amsmath}

\DeclareMathOperator*{\argmax}{arg\,max}
\DeclareMathOperator*{\argmin}{arg\,min}

\newcommand{\mr}[1]{\mathrm{#1}}
\DeclareMathOperator{\atant}{\mr{atan2}}

\newcommand\norm[1]{\left\lVert #1\right\rVert}
\newcommand\norms[1]{\big\lVert #1\big\rVert}

\newcommand{\pp}{\boldsymbol{p}}
\newcommand{\vv}{\boldsymbol{v}}

\newcommand{\qq}{\boldsymbol{q}}

\newcommand{\yy}{\boldsymbol{y}}
\newcommand{\xx}{\boldsymbol{x}}
\newcommand{\zz}{\boldsymbol{z}}
\newcommand{\sss}{\boldsymbol{s}}
\newcommand{\rr}{\boldsymbol{r}}

\newcommand{\nn}{\boldsymbol{n}}

\newcommand{\rmlos}{{\rm{LoS}}}
\newcommand{\rmnlos}{{\rm{NLoS}}}
\newcommand{\yynlos}{\yy_{\rmnlos}}
\newcommand{\yylos}{\yy_{\rmlos}}
\newcommand{\rmmpc}{{\rm{MPC}}}
\newcommand{\rmrisr}{{\rm{RISr}}}

\newcommand{\bb}{\boldsymbol{b}}

\newcommand{\bx}{\boldsymbol{x}}

\newcommand{\etab}{\boldsymbol{\eta}}

\newcommand{\bE}{\mathbf{E}}
\newcommand{\bc}{\mathbf{c}}
\newcommand{\bN}{\mathbf{N}}

\newcommand{\bH}{\mathbf{H}}
\newcommand{\bI}{\mathbf{I}}
\newcommand{\bh}{\mathbf{h}}
\newcommand{\eye}{\mathbf{I}}

\newcommand{\bn}{\mathbf{n}}
\newcommand{\FF}{\boldsymbol{F}}
\newcommand{\JJ}{\boldsymbol{J}}

\newcommand{\zetab}{\boldsymbol{\zeta}}

\newcommand{\etabch}{\etab_{\rm{ch}}}

\newcommand{\llr}{\mathcal{L}}
\newcommand{\llrlog}{\mathcal{L}^{\rm{log}}}
\newcommand{\hdet}{ \underset{\mathcal{H}_0}{\overset{\mathcal{H}_1}{\gtrless}} }

\newcommand{\cfo}{\nu}

\newcommand{\boldzero}{{ {\boldsymbol{0}} }}

\newcommand{\diag}{ \mathrm{diag}  }
\newcommand{\thn}[1]{ {#1^{\rm{th} } } }

\newcommand{\ba}{\boldsymbol{a}}
\newcommand{\bA}{\boldsymbol{A}}

\newcommand{\bY}{\mathbf{Y}}

\newcommand{\bz}{\boldsymbol{z}}
\newcommand{\bR}{\boldsymbol{R}}

\newcommand{\bW}{\boldsymbol{W}}
\newcommand{\bV}{\boldsymbol{V}}

\newcommand{\bC}{\mathbf{C}}
\newcommand{\bD}{\mathbf{D}}
\newcommand{\bP}{\boldsymbol{P}}

\newcommand{\bu}{\mathbf{u}}

\newcommand{\aab}{\boldsymbol{a}}
\newcommand{\kk}{\boldsymbol{k}}

\newcommand{\alphalos}{\alpha_0}

\newcommand{\alpharisi}{\alpha_i}
\newcommand{\alpharisr}{\alpha_r}

\newcommand{\nulos}{\nu}

\newcommand{\nurisi}{\nu_i}

\newcommand{\alpharer}{\alpha_{{\rm{R}},r}}
\newcommand{\alphaimr}{\alpha_{{\rm{I}},r}}
\newcommand{\alphare}{\alpha_{{\rm{R}},0}}
\newcommand{\alphaim}{\alpha_{{\rm{I}},0}}
\newcommand{\alphareo}{\alpha_{{\rm{R}},1}}
\newcommand{\alphaimo}{\alpha_{{\rm{I}},1}}

\newcommand{\alphareR}{\alpha_{{\rm{R}},R}}
\newcommand{\alphaimR}{\alpha_{{\rm{I}},R}}

\newcommand{\ppbs}{\pp_{\BS}}

\newcommand{\ppriso}{\pp_{{\RIS},1}}
\newcommand{\pprist}{\pp_{{\RIS},2}}
\newcommand{\pprisi}{\pp_{{\RIS},i}}
\newcommand{\pprisr}{\pp_{{\RIS},r}}

\newcommand{\vvbs}{v_{\BS}}

\newcommand{\vvrisi}{v_{{\RIS},i}}

\newcommand{\BS}{\text{BS}}
\newcommand{\RIS}{\text{RIS}}
\newcommand{\LoS}{\text{LoS}}
\newcommand{\NLoS}{\text{NLoS}}

\newcommand{\Psq}{\sqrt{P}}
\newcommand{\Ts}{T_s}

\newcommand{\complexset}[2]{ \mathbb{C}^{#1 \times #2}  }
\newcommand{\complexsetone}[1]{ \mathbb{C}^{#1}  }
\newcommand{\complexsett}{ \mathbb{C}  }
\newcommand{\realset}[2]{ \mathbb{R}^{#1 \times #2}  }
\newcommand{\realsetone}[1]{ \mathbb{R}^{#1}  }
\newcommand{\realsett}{ \mathbb{R}  }

\newcommand{\realp}[1]{ \Re \left\{#1\right\}  }
\newcommand{\imp}[1]{ \Im \left\{#1\right\}  }

\newcommand{\her}{\mathsf{H}}
\newcommand{\trp}{\mathsf{T}}

\newcommand{\gammab}{\bm{\gamma}}

\newcommand{\phib}{\bm{\phi}}
\newcommand{\psib}{\bm{\psi}}
\newcommand{\thetab}{\bm{\theta}}

\newcommand{\Chib}{\bm{\chi}}

\newcommand{\phibr}{\phib_{r}}

\newcommand{\psibaz}{[\psib]_{\rm{az}}}
\newcommand{\psibel}{[\psib]_{\rm{el}}}

\newcommand{\gammabrm}{\gammab_{r,m}}

\newcommand{\gammabrf}[1]{\gammab_{r,#1}}

\newcommand{\thetabi}{\thetab_{i}}

\newcommand{\thetabr}{\thetab_{r}}
\newcommand{\thetabraz}{[\thetabr]_{\rm{az}}}
\newcommand{\thetabrel}{[\thetabr]_{\rm{el}}}

\newcommand{\thetabio}{\thetab_{1}}

\newcommand{\thetabiR}{\thetab_{R}}

\newcommand{\hlosm}{h_{{\LoS},m}}
\newcommand{\hrism}{h_{{\RIS},m}}

\newcommand{\mtcn}{{\mathcal{CN}}}
\newcommand{\Imatrix}{{ \boldsymbol{\mathrm{I}} }}

\makeatletter \renewcommand\d[1]{\ensuremath{%
		\;\mathrm{d}#1\@ifnextchar\d{\!}{}}}
\makeatother

\makeatletter
\newcommand*\rel@kern[1]{\kern#1\dimexpr\macc@kerna}
\newcommand*\widebar[1]{%
  \begingroup
  \def\mathaccent##1##2{%
    \rel@kern{0.8}%
    \overline{\rel@kern{-0.8}\macc@nucleus\rel@kern{0.2}}%
    \rel@kern{-0.2}%
  }%
  \macc@depth\@ne
  \let\math@bgroup\@empty \let\math@egroup\macc@set@skewchar
  \mathsurround\z@ \frozen@everymath{\mathgroup\macc@group\relax}%
  \macc@set@skewchar\relax
  \let\mathaccentV\macc@nested@a
  \macc@nested@a\relax111{#1}%
  \endgroup
}
\makeatother

\newtheorem{remark}{Remark}

\acrodef{RIS}{reconfigurable intelligent surface}
\acrodef{SNR}{signal-to-noise ratio}
\acrodef{ISAC}{integrated sensing and communication}
\acrodef{ISLAC}{integrated sensing, localization, and communication}
\acrodef{LoS}{line-of-sight}
\acrodef{NLoS}{non-line-of-sight}
\acrodef{AoA}{angle-of-arrival}
\acrodef{AoD}{angle-of-departure}
\acrodef{UE}{user equipment}
\acrodef{NF}{near-field}
\acrodef{BS}{base station}
\acrodef{MCRB}{misspecified Cram\'{e}r-Rao bound}
\acrodef{CRB}{Cram\'{e}r-Rao bound}
\acrodef{LB}{lower bound}
\acrodef{ML}{maximum-likelihood}
\acrodef{MML}{mismatched maximum-likelihood}
\acrodef{DL}{downlink}
\acrodef{UL}{uplink}
\acrodef{MIMO}{multiple-input multiple-output}
\acrodef{MISO}{multiple-input single-output}
\acrodef{SISO}{single-input single-output}
\acrodef{SIP}{shift invariance property}
\acrodef{FIM}{Fisher information matrix}
\acrodef{RMSE}{root mean-squared error}
\acrodef{AWGN}{additive white Gaussian noise}
\acrodef{ADMM}{alternating direction method of multipliers}
\acrodef{LS}{least-squares}
\acrodef{SOC}{second-order cone}
\acrodef{CFO}{carrier frequency offset}
\acrodef{GLRT}{generalized likelihood ratio test}
\acrodef{FSPL}{free space path loss}
\acrodef{TDoA}{time-difference-of-arrival}
\acrodef{MPC}{multi-path components}
\acrodef{NB}{narrowband}
\acrodef{WB}{wideband}
\acrodef{TDM}{time-division multiplexing}
\acrodef{PLL}{phase-locked loop}
\acrodef{EKF}{extended Kalman filter}
\acrodef{ToA}{time-of-arrival}
\acrodef{GPS}{Global Positioning System}

\usepackage{balance}

\begin{document}
\bstctlcite{IEEEexample:BSTcontrol}

\title{Frugal RIS-aided 3D Localization with CFO under LoS and NLoS Conditions}
\author{Yasaman Ettefagh,~\IEEEmembership{Student Member,~IEEE}, Musa Furkan Keskin,~\IEEEmembership{Member,~IEEE}, Kamran Keykhosravi,~\IEEEmembership{Member,~IEEE}, Gonzalo Seco-Granados,~\IEEEmembership{Fellow,~IEEE}, and Henk Wymeersch,~\IEEEmembership{Fellow,~IEEE}
\thanks{This work is supported by the SNS JU project 6G-DISAC under the EU's Horizon Europe research and innovation Program under Grant Agreement No 101139130, in part by the Catalan Government under the ICREA Academia Program and grant 2021 SGR 00737, in part by the Spanish project PID2023-152820OB-I00 funded by MICIU/AEI/10.13039/501100011033 and ERDF/EU, and by Swedish Research council (Grant no. 2022-03007 and 2024-04390).

Yasaman Ettefagh, Musa Furkan Keskin and Henk Wymeersch are with the Department of Electrical Engineering, Chalmers University of Technology, 41296 Gothenburg, Sweden (emails: ettefagh@chalmers.se; furkan@chalmers.se; henkw@chalmers.se). 

Kamran Keykhosravi is with Ericsson Research, Ericsson AB, Gothenburg, Sweden (e-mail:kamran.keykhosravi@ericsson.com). 

Gonzalo Seco-Granados is with the Department of Telecommunications and Systems Engineering, Universitat Autonòma de Barcelona, 08193 Bellaterra,
Spain (e-mail: gonzalo.seco@uab.cat).}}

\maketitle

\begin{abstract}
In this paper, we investigate 3-D localization and frequency synchronization with multiple \acp{RIS} in the presence of \ac{CFO} for a stationary \ac{UE}. In line with the 6G goals of sustainability and efficiency, we focus on a \textit{frugal} communication scenario with minimal spatial and spectral resources (i.e., narrowband single-input single-ouput system), considering both the presence and blockage of the 
\ac{LoS} path between the \ac{BS} and the \ac{UE}. 
We design a \ac{GLRT}-based \ac{LoS} detector, channel parameter estimation and localization algorithms, with varying complexity. 
To verify the efficiency of our estimators, we compare the \ac{RMSE} to the \ac{CRB} of the unknown parameters. 
We also evaluate the sensitivity of our algorithms to the presence of uncontrolled \ac{MPC} and various levels of \ac{CFO}. Simulation results showcase the effectiveness of the proposed algorithms under minimal hardware and spectral requirements, and a wide range of operating conditions, thereby confirming the viability of RIS-aided frugal localization in 6G scenarios.  
\end{abstract}



\acresetall 
\section{Introduction}\label{sec:intro}
\IEEEPARstart{A}{ccurate} positioning is a pre-requisite for many modern use-cases, finding applications in various areas such as autonomous driving, augmented reality, navigation, etc. \cite{c542dde194c4451abf82fb0cf3c10301}. 
While \ac{GPS} stands out as the ubiquitous solution for navigation, its efficacy is often compromised in scenarios where a direct \ac{LoS} to satellites is obstructed, such as in tunnels and dense urban environments.
An alternative solution is to use cellular networks. Cellular networks have been used for positioning since the first generation (1G) of mobile networks \cite{del2017survey}. 
In 4G networks, \ac{TDoA} and \ac{AoA}-based techniques were introduced, which rely on multiple base stations to estimate the position of a mobile device. These techniques proved to be more accurate than the previous ones, but required additional infrastructure. In 5G networks, the adoption of mmWave frequencies and beamforming techniques improved the accuracy of positioning 
\cite{italiano2023tutorial,wen2019survey,shahmansoori2017position}. In 6G networks, the use of \acp{RIS} is expected to further enhance the accuracy and reliability of cellular-based localization \cite{wymeersch2020radio, behravan2022positioning, emenonye2023fundamentals, he2022beyond}. 

In advancement of wireless communications through successive generations, achieving sustainability and improving energy and spectral efficiency has always been a central objective, and the upcoming 6G is no exception \cite{tataria20216g}. As pointed out, 4G networks succeeded in improving positioning by putting a minimum requirement on hardware resources and 5G managed to progress further in positioning by accessing large chunks of bandwidth in mmWave and large antenna arrays. Therefore, the question remains whether 6G will continue this trend or if there could be another solution to achieve improved positioning without leveraging more resources. The use of \acp{RIS} appears to be a key point in this regard.

 \acp{RIS} constitute one of the key-technology enablers for 6G \cite{RIS_Access_2019}, providing an additional, controllable path between the \ac{BS} and the \ac{UE}. Primarily, \acp{RIS} are designed as a cost-effective and energy-efficient solution to address signal blockage challenges without the need to densify the network with more \acp{BS}. This approach is favored due to the simpler hardware requirements and lower maintenance costs associated with \ac{RIS} \cite{pan2021reconfigurable}, \cite{jian2022reconfigurable}. 
\acp{RIS} consist of a large number of small elements that can be manipulated to reflect incident waves in desired directions \cite{RIS_tutorial_2021}. Deploying \acp{RIS} in an environment allows for the engineering of the propagation medium, enhancing signal strength and reducing interference at targeted locations \cite{pan2022overview}. 
This ultimately enhances communication rate and coverage while providing significant gains in localization performance \cite{bjornson2021reconfigurable}.

The use of \acp{RIS} in radio localization has been extensively studied in the literature, with numerous works exploring the potential of \acp{RIS} to improve \ac{UE} localization thanks to the additional reflected paths \cite{de2021convergent,umer2023role, li2024star, sun2024computational, JSTSP_RIS_2022}. Notable contributions include \cite{wymeersch2020radio,de2021convergent,umer2023role}, which discuss the challenges, opportunities, and research directions related to \ac{RIS}-aided positioning. Since localization often comes as a by-product of cellular communication systems, it is advantageous if it is \textit{frugal} with resources, such as antennas and time-frequency allocations, to avoid compromising the quality and infrastructure requirements of wireless communications—so far the primary objective of cellular networks.
Along this line, 
\ac{RIS}-enabled localization with single \ac{BS} and single antenna is discussed in \cite{Kamran_JSTSP_SISO_RIS,ye2022single}. In \cite{Kamran_JSTSP_SISO_RIS}, a localization and clock-synchronization approach to \ac{SISO}, single \ac{RIS} \ac{WB} case was introduced. In \cite{ye2022single}, authors propose a method for single-antenna receiver 2-D localization and achieve centimeter-level localization accuracy with fingerprinting the \ac{RIS}-phases over time.
A broader perspective  on frugal localization was taken in  \cite{keykhosravi2023leveraging}, which conducted an analysis on \ac{UE} localization scenarios with minimal required number of \ac{BS} and \acp{RIS}, showing that it is possible to estimate the \ac{UE}'s position with one \ac{BS} and two \acp{RIS} under \ac{NB} communication even without a direct path from the \ac{BS} to the \ac{UE}. 

A common weakness in the above works is that they ignore the presence of \ac{CFO} between the \ac{BS} and the \ac{UE}. Since both \ac{CFO} and the use of diverse RIS phase profiles \cite{keykhosravi2021siso} (i.e., beam sweeping for \ac{AoD} estimation from the \ac{RIS} using beamspace measurements at the \ac{UE} \cite{beamspace_loc_2020,JSTSP_RIS_2022}) over time lead to phase variations across consecutive transmissions, the impacts of \ac{CFO} and RIS profile variation add up together in the time-series of phase shifts, resulting in the so-called \textit{angle-Doppler/CFO coupling effect}, as noted in \cite{tdm_mimo_radar_JSTSP_2021,ercan2024ris}. 
 In real systems, accurate estimation and compensation of CFO are vital for localization and communication performance. Techniques vary by application; for instance, in \cite{gong2021joint} a pilot-based approach using orthogonal pilots and a Hadamard matrix structure allows efficient CFO estimation and compensation, enhancing localization accuracy by reducing phase distortions. Similarly, \cite{yao2011joint} employs a synthetic aperture method to jointly estimate AoD and CFO, improving resolution in multi-path environments. CFO estimation and compensation are also covered extensively in standard communication literature, such as \cite[Ch.~6~\&~8]{meyr1998digital}, where both pilot-based and non-data-aided techniques are discussed as effective solutions for mitigating CFO in practical systems. 
From an experimental perspective, \cite{keykhosravi2023leveraging} demonstrates \ac{SISO} localization of a stationary \ac{UE} with one \ac{BS} and two \acp{RIS} in the absence of a \ac{CFO}. However, this study has the impractical requirement that the BS and UE share a common oscillator to eliminate CFO issues. Moreover, the approach relies on using directional RIS beams to sweep potential locations and estimate AoDs from the RISs. This method is ineffective with random RIS phase profiles, which are commonly used in RIS-aided communications when UE locations are unknown, e.g., \cite{RIS_bounds_TSP_2021,ris_nakagami,ris_mimo_phase_2022}.
Overall, no systematic study has been conducted to address the problem of \textit{frugal localization} and frequency synchronization with the help of \acp{RIS} under angle/CFO coupling effect and when employing random \ac{RIS} phase configurations. Therefore, in light of the existing literature, two crucial questions emerge that remain unanswered: \textit{(i)} is it possible to perform joint 3-D downlink localization and frequency synchronization in the challenging \ac{NB} (i.e., single-carrier) \ac{SISO} scenario with multiple \acp{RIS} employing random phase configurations?; and \textit{(ii)} can efficient and low-complexity algorithms be developed for \ac{LoS} presence detection, channel parameter/CFO estimation and localization in both the presence and absence of the direct \ac{LoS} path between the \ac{BS} and the \ac{UE}?


In an attempt to address the identified shortcomings of existing studies on \ac{RIS}-aided localization and fill the corresponding research gaps, we consider the frugal localization problem of a \ac{UE} in case of \ac{NB} \ac{SISO} signaling via one \ac{BS} and several \acp{RIS} employing random phase profiles. 
The novel contributions of this paper can be summarized as follows.
\begin{itemize}
    \item \textbf{Frugal localization under CFO:} We formalize the problem of \ac{NB} RIS-aided SISO localization and frequency synchronization under the assumption of \ac{CFO} between the transmitter and the receiver with parsimonious usage of resources.
     We develop novel low-complexity estimators of the unknowns, including the \ac{CFO} and \acp{AoD}, which enable us to perform localization. Different estimation algorithms under various circumstances are designed, including one estimation algorithm in case the direct \ac{LoS} between the \ac{BS} and the \ac{UE} is present, as well as two estimation algorithms in case of blocked \ac{LoS} between the \ac{BS} and \ac{UE}. Our estimators achieve the theoretical bounds at moderate to high transmit power. Our estimators are inspired by \ac{ML} estimation, but unlike conventional ML approaches—which suffer from high computational complexity due to high-dimensional search spaces—our estimators maintain very low complexity while still achieving the theoretical bounds at moderate to high transmit powers. This balance between accuracy and efficiency constitutes the core novelty of our approach. Simulation results show the efficiency of the estimator with respect to theoretical bounds.
    \item \textbf{LoS detection:} We design a \ac{GLRT}-based detector to determine if the \ac{LoS} exists or not, which complements the above contribution to a full localization and synchronization algorithm. Our proposed algorithm is capable of attaining the theoretical bounds at a moderate to high transmit power.
    \item \textbf{Sensitivity analysis:} We assess the sensitivity of our algorithm to \ac{MPC}, which shows that our algorithm quickly converges to the achievable bounds with Rician factor as small as $10$.
    We also study the sensitivity of location estimation under UE motion. Moreover, we compare the performance of our algorithm with prior works in which CFO is not compensated for by ignoring
    \ac{CFO} as an unknown. The performance drastically deteriorates at \ac{CFO} values much smaller than values typically found in \ac{UE}, suggesting the importance of accurate estimation of the \ac{CFO}. The analyses provide a detailed insight about the applicability of the proposed algorithms.
    \item \textbf{Resource Efficiency:} We prove that with a minimalistic spectral and BS/UE hardware configuration, including a single antenna and single subcarrier and sufficient number of \acp{RIS}, it is possible to perform accurate positioning. Therefore, our approach sets a new benchmark for resource efficiency in communication systems, potentially reducing the cost and complexity of deployment, which is in accordance with 6G objectives.
\end{itemize}

The paper is structured as follows. In Sec.~\ref{sec:sysmod}, the system setup is detailed. Sec.~\ref{subsec:RIS_PD} describes orthogonal RIS profile design to facilitate per-RIS \ac{AoD} and \ac{CFO} estimation. 
Sec.~\ref{sec:highLevel} describes the overall flow of the algorithm, including channel parameter estimation under \ac{LoS} and \ac{NLoS}, localization  under \ac{LoS} and \ac{NLoS}, and detection of if the \ac{LoS} is present or not. Channel parameter turns out to be the main challenge and low-complexity methods are derived in Sec.~\ref{sec:ChParamEst}: 
in Sec.~\ref{subsec:LoS_localization}, estimation algorithms in case the direct link between the \ac{BS} and \ac{UE} (\ac{LoS}) exists will be explained. Sec.~\ref{subsec:NLoS_localization} sheds light on estimation algorithms in the more challenging scenario in which the \ac{LoS} path does not exist. 
The performance of the estimators and detector are shown and discussed in Sec.~\ref{sec:results}. Finally, Sec.~\ref{sec:conclusion} concludes the findings of the paper. 

\subsubsection*{Notation}
Vectors and matrices are shown by bold-face lower-case and bold-face upper-case letters respectively. The notations $(.)^\trp$ and $(.)^\her$ indicate transpose and hermitian transpose. All one vector with size $n$ denoted by $\boldsymbol{1}_n$ and the identity matrix with size $n$ is represented by $\eye_n$. The L2 norm of a vector is shown by $\norm{.}$. The matrix $\bR_z(\theta)$ denotes the 3-D rotation matrix by an angle $\theta$ about the z-axis. The Kronecker product and the Hadamard product are shown by $\otimes$ and $\odot$ respectively.

\section{System Model}\label{sec:sysmod}
In this section, we describe the proposed localization system that promotes the parsimonious use of spatial and spectral resources for joint location and frequency offset estimation of a static user.


\subsection{Scenario}
\begin{figure}[t]
    \hspace{-0.15cm}
    \includegraphics[width = 0.45\textwidth]{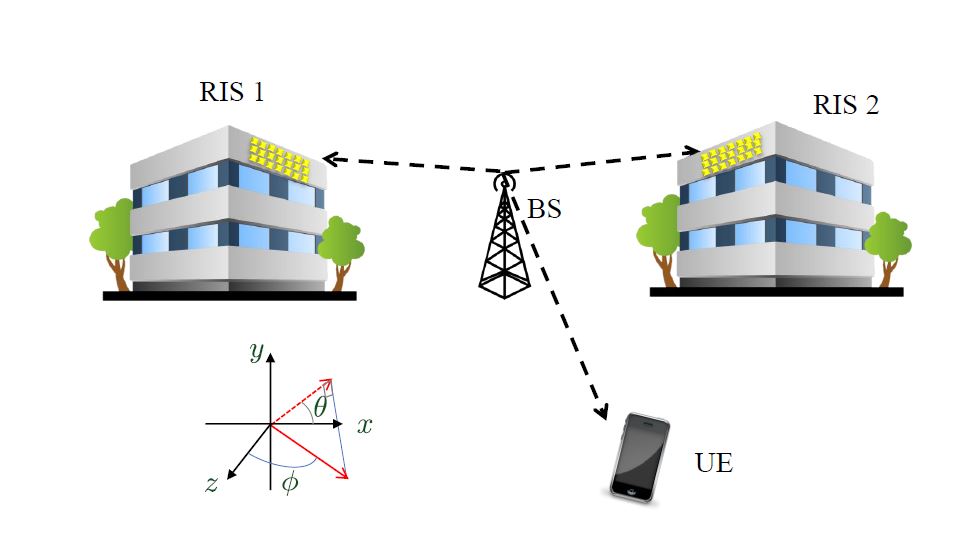}
    \caption{Frugal localization scenario involving one single-antenna \ac{BS}, $R = 2$ \acp{RIS} and one single-antenna user with \ac{NB} (single-carrier) communication.}
    \label{fig:scenarioRIS}
\end{figure}

Consider a RIS-aided \ac{DL} localization system with a single-antenna \ac{BS}, $R$ identical $N$-element \acp{RIS}, and a single-antenna \ac{UE}, as shown in Fig.~\ref{fig:scenarioRIS}. The \ac{BS} and the \acp{RIS} are located at known positions $\ppbs \in \realsetone{3}$ and $\pprisr \in \realsetone{3}$, $r = 1, \cdots, R$ (denoting the \ac{RIS} centers) respectively, while the \ac{UE} has an unknown position $\pp \in \realsetone{3}$. The \acp{RIS} are assumed to have known orientations, represented by the unitary rotation matrices $\bR_r  \in \text{SO}(3) \subset \realset{3}{3}$, $r = 1, \cdots, R$, that map the global frame of reference to the local coordinate systems of the \acp{RIS}. The \ac{UE} is considered stationary. Moreover, due to oscillator inaccuracies, the \ac{UE} is not perfectly frequency-synchronized to the \ac{BS}, which leads to an unknown \ac{CFO} $\cfo \in \realsett$ at the \ac{UE} with respect to the \ac{BS} \cite{Visa_CFO_TSP_2006,CFO_TVT_2011}.


\subsection{Geometric Relations}
The \ac{AoD} from the $\thn{r}$ \ac{RIS} to the \ac{UE} is denoted by $\thetabr = \big[\thetabraz, \thetabrel \big]^\trp \in \realsetone{2}$, where
\begin{align} \label{eq:geo_rel1}
    \thetabraz &= \atant\left( [\rr_r]_2, [\rr_r]_1 \right) ,
    \\
    \thetabrel &=\arccos \left(\frac{[\rr_r]_3}{\norm{\pp-\pprisr}}\right) , \label{eq:geo_rel2}
\end{align}
with $\rr_r = \bR_r (\pp - \pprisr)$ representing the vector extending from the $\thn{r}$ \ac{RIS} to the \ac{UE} in the local frame of reference of the $\thn{r}$ \ac{RIS}.


\subsection{Signal Model}
The \ac{BS} transmits \ac{NB} pilot symbols $\sss = [s_0 \, \cdots \, s_{M-1}]^\trp \in \complexsetone{M}$ over $M$ transmission instances with sampling period $\Ts$ and a power constraint $P$ such that $\norm{\sss}^2 = MP$. We denote the continuous-time transmit signal as $s(t) = \sum_{m=0}^{M-1} s_m q(t-m\Ts)$, where $q(t)$ is any unit-energy pulse. Assuming the absence of uncontrolled multipath 
(i.e., due to reflection or scattering off passive objects) as in \cite{Kamran_JSTSP_SISO_RIS,RIS_bounds_TSP_2021,zhang2020towards,RIS_Location_Win_2022,JSTSP_RIS_2022}, the received complex baseband signal at the \ac{UE}, corresponding to the $\thn{m}$ transmission and after matched filtering \footnote{The transmit pulse $q(t)$ is deigned such that the resulting effective pulse $g(t) = q(t) \circledast q^*(-t)$ satisfies the Nyquist criterion.}, is given by
\begin{align}\label{eq:ym}
    y_m = h_m e^{j 2 \pi m \Ts \cfo} s_m + n_m ,
\end{align}
where $\Ts \in \realsett$ is the symbol duration, the term $e^{j 2 \pi m \Ts \cfo}$ results from the \ac{CFO} ($\nu$) between the \ac{BS} and the \ac{UE}, $n_m \in \complexsett$ denotes circularly symmetric complex Gaussian noise with $n_m \sim \mtcn(0, \sigma^2)$, and $h_m \in \complexsett$ represents the overall \ac{BS}-\ac{UE} channel for the $\thn{m}$ transmission involving both the \ac{LoS} path and the \ac{NLoS} paths through the \acp{RIS}, i.e.,\footnote{In a section of the simulation results (Sec.~\ref{subsec:result_MPC}), uncontrolled \ac{MPC} will be incorporated into the channel model as part of a sensitivity analysis.}
\begin{align}\label{eq_hm}
    h_m = \hlosm + \hrism.
\end{align}

In \eqref{eq_hm}, $\hlosm \in \complexsett$ is the \ac{LoS} (i.e., direct) channel between the \ac{BS} and the \ac{UE}, written as
\begin{align}\label{eq_hlosm}
    \hlosm = \alphalos,
\end{align}
where $\alphalos \in \complexsett$ denotes the LoS channel gain. As to the \ac{NLoS} channel in \eqref{eq_hm}, it can be defined as
\begin{align}\label{eq_hrism}
    \hrism = \sum_{r=1}^{R} \alpharisr \, \aab^\trp(\thetabr) \diag(\gammabrm) \aab(\phibr)  ,
\end{align}
where $\alpharisr \in \complexsett$ is the channel gain over the \ac{BS}-($\thn{r}$ RIS)-\ac{UE} path and $\phibr \in \realsetone{2}$ denotes the known \ac{AoA} from the \ac{BS} at the $\thn{r}$ \ac{RIS} (given the known positions and orientations of the \ac{BS} and the \acp{RIS}). In addition, $\gammabrm \in \complexsetone{N}$ represents the phase profile of the $\thn{r}$ \ac{RIS} at time $m$ and  $\aab(\cdot) \in \complexsetone{N}$ is the \ac{RIS} steering vector, given by \cite{Kamran_JSTSP_SISO_RIS}
\begin{align} 
    [\aab(\psib)]_n = \exp\left( j \kk^\trp(\psib) \qq_n \right),
\end{align}
for a generic $\psib$  where $\qq_n \in \realsetone{3}$ denotes the known position of the $\thn{n}$ \ac{RIS} element with respect to the \ac{RIS} center in the local coordinate system of the \ac{RIS}, and
\begin{align}
   &\kk(\psib) =  \frac{2\pi}{\lambda}  \\ & \left[ \sin(\psibel) \cos(\psibaz), \, \sin(\psibel) \sin(\psibaz), \, \cos(\psibel) \right]^\trp \nonumber
\end{align}
is the wavenumber vector defined for a given angle $\psib$. 
Let
\begin{align}
    \bW_r &\triangleq [ \aab(\phibr) \odot \gammabrf{0} ~ \cdots ~ \aab(\phibr) \odot \gammabrf{M-1} ] \in \complexset{N}{M} , \label{eq_bW}
    \\ \label{eq_bbnu}
    \bb(\nu) &\triangleq [ 1 ~ e^{j 2\pi \Ts \nu} ~ \cdots ~ e^{j 2\pi (M-1) \Ts \nu} ]^\trp   \in \complexsetone{M}.
\end{align}
Then, using \eqref{eq_hm}--\eqref{eq_hrism} and \eqref{eq_bW}--\eqref{eq_bbnu}, the aggregated observations in \eqref{eq:ym} over $M$ transmissions can be written as \footnote{In Sec. \ref{subsec_velocity}, we extend the received signal model to account for UE motion with velocity $\vv$. When the condition $\lvert M \Ts(\pp_i - \pp)^\trp \vv/(\lambda\norm{\pp_i - \pp})\rvert < 1/8$ holds (where $i\in \BS, R_1, ..., R_R$), the Doppler-induced phase variation over the $M$ transmissions is negligible \cite{richards2005fundamentals}.} 
\begin{align}\label{eq_yy}
    \yy = \Psq \Big( \alphalos \, \bb(\nulos) + \sum_{r=1}^{R} \alpharisr \, \bx_r(\thetabr) \odot \bb(\nulos)  \Big) + \nn ,
\end{align}
where $\yy \triangleq [y_0 \, \cdots \, y_{M-1}]^\trp \in \complexsetone{M}$, $\bx_r(\thetabr) = \bW_r^\trp \aab(\thetabr)$ and $\nn \in \complexsetone{M}$ is the noise component with $\nn \sim \mtcn(\boldzero, \sigma^2 \Imatrix)$. In \eqref{eq_yy}, we have set $s_m = \Psq, \, \forall m$ for simplicity.



\subsection{Joint Localization and Frequency Synchronization Problem}\label{sec_prob_desc}
The goal is to estimate the position $\pp$ and the \ac{CFO} $\cfo$ of the \ac{UE} from the observation $\yy$ in \eqref{eq_yy}. For this estimation problem, the unknown channel-domain parameters are given by
\begin{align} \label{nu_ch_withLoS}
    &\etabch^\LoS = [\alphare ~ \alphaim   ~ \cdots ~ \alphareR ~ \alphaimR ~ \nulos ~ \thetabio^\trp  ~ \cdots  ~ \thetabiR^\trp] \in \realsetone{4R+3}, 
\end{align}
while the unknown location-domain parameter vector is
\begin{align} \label{nu_loc_withLoS}
    \etab^\LoS = [\alphare ~ \alphaim ~ \cdots ~ \alphareR ~ \alphaimR  ~ \cfo ~ \pp^\trp]^\trp \in \realsetone{2R+6}  .
\end{align}
Here, $\alpharer \triangleq \realp{\alpha_r}$ and $\alphaimr \triangleq \imp{\alpha_r}$, $r = 0, \cdots, R$. The superscript '$\LoS$' is used to highlights that the \ac{LoS} path exists. Clearly, in case no \ac{LoS} exists between the \ac{UE} and the \ac{BS}, the channel-domain vector and the location-domain vector would change into
\begin{math} \label{nu_ch_noLoS}
    \etabch^\NLoS = [\alphareo ~ \alphaimo ~ \cdots ~ \alphareR ~ \alphaimR ~ \nulos ~ \thetabio^\trp  ~ \cdots  ~ \thetabiR^\trp] \in \realsetone{4R+1},
\end{math}
and 
\begin{math}  \label{nu_loc_noLoS}
    \etab^\NLoS = [\alphareo ~ \alphaimo ~ \cdots ~ \alphareR ~ \alphaimR  ~ \cfo ~ \pp^\trp]^\trp \in \realsetone{2R+4},
\end{math}
respectively. The superscripts '$\NLoS$' is used to denote that the \ac{LoS} path does not exist.

It should be noted that the \ac{LoS} path does not directly convey  positional information because the \ac{UE}'s position is a function of the \acp{AoD} according to \eqref{eq:geo_rel1}--\eqref{eq:geo_rel2}. In our formulation, the LoS component appears solely as a complex scalar whose phase evolves over time according to the CFO, and does not include any AoD-dependent terms. However, as we will see in Sec.~\ref{sec:results}, the accuracy of \ac{CFO} estimation is improved if the \ac{LoS} is present, and as a result, the residual error in \ac{RIS} path separation\footnote{To be able to estimate \acp{AoD} separately from each \ac{RIS}, signal components corresponding to different \acp{RIS} in \eqref{eq_yy} need to be separated.} decreases as discussed in Sec.~\ref{subsec:RIS_PD}, simplifying also the \ac{AoD} estimation algorithm.




\section{\ac{RIS} Profile Design} \label{subsec:RIS_PD}
Estimating the channel-domain parameters in \eqref{nu_ch_withLoS} requires a complex high-dimensional optimization, which is cumbersome even in the case of $R=2$. To circumvent this, we leverage the controllability offered by the RIS by 
designing an orthogonal temporal-coding for \ac{RIS} phase profiles \cite{keykhosravi2021multiris}. In the absence of \ac{CFO}, the contributions from $R+1$ paths in \eqref{eq_yy} can be separated and the unknowns from each \ac{RIS} path can be estimated separately.

\subsection{Hadamard-based Design}
The idea involves using the rows of the Hadamard matrix
to encode the phase profiles of the \acp{RIS}, followed by a simple post processing at the receiver to retrieve the contributions from each path \cite{keykhosravi2021multiris}. 

We first 
    divide the total transmissions into $L \geq 2^{\lceil{\log_2(R+1)}\rceil}$ equal-sized blocks. Choose $L$ such that it is a factor of $M$.
We next define $\bP_r \in \mathbb{C}^{N \times (M/L)}$, $r = 1,\cdots, R$ as a  set of \emph{base phase profiles} of length $M/L$ for each \ac{RIS}. These profiles may be random in case there is no prior information about the \ac{UE} location, or directional in case partial information about the \ac{UE} location is available.

    We take $R$ rows of the Hadamard matrix\footnote{Recall that a Hadamard matrix $\bC$ of length $L$ is an $L\times L$ matrix, is made up of entries in $\{ -1, +1\}$, and satisfies $\bC \bC^\top = L \bI$. The amplitude constraint is compatible with the RIS profile constraint. } of length $L$ (except for the first row, which is constant) as the coding vectors $\bc_r \in \realsetone{L}$ for the $\thn{r}$ \ac{RIS}, $r = 1,\cdots, R$. Then we then form the full phase profile  of each \ac{RIS} at transmission $m=kL+l$,  $k = 0, \cdots, M/L-1$,
    \begin{align} \label{eq:gammaRISPP}
        \boldsymbol{\gamma}_{r, kL+l} = \left([\bc_r]_{l}\right) {\bP_r}_{[:,k]}, 
    \end{align} %
    where ${\bP_r}_{[:,k]}$ denotes the $\thn{k}$ column of $\bP_r$, $[\bc_r]_{l}$ denotes the $\thn{l}$ element of $\bc_r$, and $l = 0, \cdots, L-1$. This indexing structure is used to map the two indices $(k, l)$ into a single index $m$, where $1 \leq m \leq M$. 

At the receiver side, 
    to extract the $\thn{r}$ \ac{RIS} components from \eqref{eq_yy}, it is enough to reshape the received signal $\yy \in \mathbb{C}^{M}$ as
    \begin{align} \label{eq:YmatReshape}
    \bY = [\yy_{0:L-1}| \yy_{L:2L-1} | \cdots|\yy_{M-L:M-1} ] \in \complexset{L}{(M/L)},
\end{align}
    and then compute 
    \begin{align} \label{eq:time_coding}
        \yy_{r} = \frac{1}{L} \bY^\trp \bc_r, ~ r = 1,\cdots,R.
    \end{align}
    where $\yy_r \in \complexsetone{M/L}$ is the filtered signal containing the $\thn{r}$ \ac{RIS} components.
    It is possible to separate the \ac{LoS} component using
    \begin{math}
        \yy_0 = {1}/{L} \bY^\trp \bc_0,
    \end{math}
    where $\bc_0 = \boldsymbol{1}_L$. As a result, despite the RIS-reflected paths having significantly lower energy than the LoS path due to the approximately random phase configuration, the use of orthogonal temporal coding for RIS phase profiles ensures their extraction through coherent integration. The details of this extraction are provided in the next subsection, with an example.

\subsection{Example}\label{sec_example}
As an example, we fix the number of \acp{RIS} to $R=2$. Note that to separate the contributions from the three paths (\ac{LoS}, \ac{RIS} 1 and \ac{RIS} 2), the minimum required coding length is $L = 4$.
We take  $\bc_0 = [1, 1, 1, 1]^\trp$, $\bc_1 = [1, -1, 1, -1]^\trp$ and $\bc_2 = [1, 1, -1, -1]^\trp$. The vectors $\bz_0=\Psq \alphalos  \, \bb(\nulos)$ and $\bz_r =  \Psq\alpharisr \, \bx_r(\thetabr) \odot \bb(\nulos)$, $r = 1,2$ and $l = 0,1,2,3$ represent the noise-free received signal contributions from the LoS and RIS paths, respectively. These vectors capture the signal components corresponding to each path before noise is introduced.  

We can express the $4$ consecutive noise-free samples of each path as follows:  
\begin{align}
&   [\zz_0]_{4k+l} = \Psq \alpha_0 e^{j2\pi(4k+l)\Ts\nu}, \\
  &  [\zz_1]_{4k+l} = \Psq \alpha_1 \left([\bc_r]_l~g_1(\phib_1, \thetab_1, k)\right) e^{j2\pi(4k+l)\Ts\nu},
\end{align}
\begin{align}
    &[\zz_2]_{4k+l}  = \Psq \alpha_2\left([\bc_2]_l~g_2(\phib_2, \thetab_2, k)\right) e^{j2\pi(4k+l)\Ts\nu},
\end{align}
where $g_r(\phibr, \thetabr, k) = (\aab(\phi_r)\odot {\bP_r}_{[:,k]})^\trp\aab(\thetabr)$, $r=1,2$. Then, by applying \eqref{eq:time_coding} and in the case of $\nu = 0$, the filtered measurements $\yy_{0}$, $\yy_{1}$ and $\yy_{2}$ will be as follows
\begin{align} \label{yyr_0_nu}
    &[\yy_{0}]_{k} = {\Psq} \alphalos +[\tilde\nn_0]_k,
\\
    &[\yy_{1}]_{k} = {\Psq} \alpha_1 g_1(\phib_1, \thetab_1, k) +[\tilde\nn_1]_k,
\\
    &[\yy_{2}]_{k} = {\Psq} \alpha_2 g_2(\phib_2, \thetab_2, k) +[\tilde\nn_2]_k,
\end{align}
where $[\tilde\nn_r]_k = {1}/{4} \sum_{l=0}^{3}[\bc_r]_l[\nn]_{4k+l}$ (see App.~\ref{app:RIS}). Therefore, $\yy_r$, $r = 1,2$ contain the contribution from the $\thn{r}$ \ac{RIS} and $\yy_0$ contain the \ac{LoS} contribution without any interference from the other \acp{RIS}. However, if $\nu \neq 0$, this coding would result in some residual inter-RIS interference and the separation cannot be done perfectly. The solution is to first estimate the \ac{CFO} ($\hat \nu$) and remove it from the observations by $\tilde\yy = \yy \odot \bb(-\hat\nu)$, then proceed with this coding. Hence in the following sections, we proceed with estimating the \ac{CFO} and cancel it out from the received signal.



\section{High-level Algorithm Description} \label{sec:highLevel}
In this section, we provide a high-level description of the proposed joint localization and synchronization algorithm to tackle the problem in Sec.~\ref{sec_prob_desc}, 
covering both the channel-domain parameter estimation and localization. In Sec. \ref{subsec:ChParamEst}, the maximum likelihood solution to the channel parameter estimation problem is presented under both LoS and NLoS scenarios. Then, in Sec. \ref{sec_loc}, the maximum likelihood solutions to the localization and synchronization problem are presented, assuming that we already know whether the LoS link exists. Finally, Sec. \ref{sec:det} concludes the section by addressing the joint LoS detection and localization problem, meaning that the presence of the LoS path is unknown. The overall high-level algorithm is summarized in Algorithm \ref{alg:det_GLRTinsp-HW}.
\subsection{Channel-domain Parameter Estimation} \label{subsec:ChParamEst}
The \ac{ML} estimate of
the channel-domain parameters from \eqref{eq_yy} can be obtained by solving
the following optimization problems:
\begin{align} \label{eq:ML_LoS_RIS}
     &[\hat\nu, \hat\thetab_1, \cdots, \hat\thetab_R, \hat\alphalos, \hat\alpha_1, \cdots, \hat\alpha_R] =  \\ & \argmin_{\substack{\nu, \thetabio, \cdots, \thetabiR, \\ \alphalos, \alpha_1, \cdots, \alpha_R}} \norms{\yy - \Psq \alpha_0\bb(\nu) - \Psq\sum_{r=1}^{R}\alpha_r\xx_r(\thetabr)\odot\bb(\nu)}^2 \nonumber \,,
\end{align}
if \ac{LoS} exists, which we will refer to as LoS scenario, and 
\begin{align} \label{eq:ML_NLoS_RIS}
     &[\hat\nu, \hat\thetab_1, \cdots, \hat\thetab_R, \hat\alpha_1, \cdots, \hat\alpha_R] =  \\ &\argmin_{\substack{\nu, \thetabio, \cdots, \thetabiR, \\ \alpha_1, \cdots, \alpha_R}} \norms{\yy -\Psq\sum_{r=1}^{R}\alpha_r\xx_r(\thetabr)\odot\bb(\nu)}^2 \nonumber \,,
\end{align}
if \ac{LoS} is blocked, which we will refer to as NLoS scenario. 

\subsubsection{Conditional Channel Gain Estimation}
The \ac{ML} estimate of the path gains can be derived easily in closed-form as a function of the estimated \ac{CFO} and \acp{AoD} as follows. 
First, we start by writing the received signal in the below form
\begin{align} 
    \yy = \bA(\Chib_\text{ch}) \boldsymbol{\alpha} + \nn,
\end{align}
where $\Chib_\text{ch} = [\nu, \thetab_1^\trp, \cdots, \thetab_R^\trp]^\trp$, $\bA(\Chib_\text{ch})  = \Psq[\bb(\nu) ~~ \xx_1(\thetab_1) \odot \bb(\nu) ~~ \cdots ~~ \xx_R(\thetab_R) \odot \bb(\nu)]$ and $\boldsymbol{\alpha} = [ \alphalos~\alpha_1~ \cdots ~\alpha_R]^\trp,$
in case \ac{LoS} exists and $\bA(\Chib_\text{ch})   = \Psq[\xx_1(\thetab_1) \odot \bb(\nu) ~~ \cdots ~~\xx_R(\thetab_R) \odot \bb(\nu)]$ and $\boldsymbol{\alpha} = [ \alpha_1 ~\cdots ~ \alpha_R]^\trp$
in case \ac{LoS} is obstructed. Accordingly, the path gains can be estimated in closed-form as 
\begin{align} \label{eq:pathGainEstimate}
    \hat{\boldsymbol{\alpha}}(\Chib_\text{ch})  = (\bA(\Chib_\text{ch}) ^\her\bA(\Chib_\text{ch}) )^{-1}\bA(\Chib_\text{ch}) ^\her\yy.
\end{align}


\subsubsection{Compressed Channel Parameter Estimation}
We can now plug the channel gain estimates in \eqref{eq:pathGainEstimate} back into \eqref{eq:ML_LoS_RIS} and \eqref{eq:ML_NLoS_RIS} to derive the compressed \ac{ML} cost functions, reducing their dimensionality to $(2R+1)$ (which, however, still lead to very high computational complexity):
 \begin{align} \label{eq:LoS_formulation}
    &[\hat\nu, \hat\thetab_1, \cdots, \hat\thetab_R] = \argmin_{\nu, \thetabio, \cdots, \thetabiR} \\ &\norms{\yy -  \Psq \hat\alpha_0(\Chib_\text{ch})\bb(\nu) - \Psq\sum_{r=1}^{R}\hat\alpha_r(\Chib_\text{ch})\xx_r(\thetabr)\odot\bb(\nu)}^2. \nonumber
 \end{align}
in case \ac{LoS} exists, and
\begin{align} \label{eq:NLoS_formulation}
    &[\hat\nu, \hat\thetab_1, \cdots, \hat\thetab_R] = \\ &\argmin_{\nu, \thetabio, \cdots, \thetabiR} \norms{\yy -\Psq\sum_{r=1}^{R}\hat\alpha_r(\Chib_\text{ch})\xx_r(\thetabr)\odot\bb(\nu)}^2.\nonumber
 \end{align}
in case \ac{LoS} is obstructed. 

As discussed in Sec.~\ref{subsec:RIS_PD}, it is possible to significantly simplify the optimization problems \eqref{eq:LoS_formulation} and \eqref{eq:NLoS_formulation} with time-orthogonal \ac{RIS} phase profiles, assuming a good estimate of the  \ac{CFO} is available. 
Sec.~\ref{sec:ChParamEst} will provide a detailed explanation of how to solve these two optimization problems with reasonable complexity, leveraging the orthogonal RIS profile design.

\subsection{Localization}\label{sec_loc}
The direct \ac{ML} approach to solve the localization and synchronization problem in in Sec.~\ref{sec_prob_desc} is as follows:
\begin{align} \label{eq:opt_localization}
     &[\hat\nu, ~ \hat\pp] =\argmin_{\nu, \pp} \norms{ \yy -\nonumber \\ &\Psq \hat\alpha_0(\Chib_\text{p})\bb(\nu) - \Psq\sum_{r=1}^{R}\hat\alpha_r(\Chib_\text{p})\xx_r(\pp)\odot\bb(\nu)}^2,
\end{align}
in LoS scenario, and
\begin{align} \label{eq:opt_localizationnLoS}
     &[\hat\nu, ~ \hat\pp] = \argmin_{\nu, \pp} \norms{\yy - \Psq\sum_{r=1}^{R}\hat\alpha_r(\Chib_\text{p})\xx_r(\pp)\odot\bb(\nu)}^2,
\end{align}
in NLoS scenario, where $\hat\alpha_0(\Chib_\text{p})$ and $\hat\alpha_r(\Chib_\text{p})$ can be found from \eqref{eq:pathGainEstimate} by replacing $\Chib_\text{p} = [\nu, \pp^\trp]$ with $\Chib_\text{ch}$ using $\bA(\Chib_\text{p})  = \Psq[\bb(\nu) ~~ \xx_1(\thetab_1(\pp)) \odot \bb(\nu) ~~ \cdots ~~ \xx_R(\thetab_R(\pp)) \odot \bb(\nu)]$, \eqref{eq:geo_rel1} and \eqref{eq:geo_rel2}.\\
Both problems can be solved using a gradient descent method, starting from an initial estimate of $\nu$ and $\pp$. The initial estimate of $\nu$ is provided directly by the channel parameter estimator in \eqref{eq:LoS_formulation} or \eqref{eq:NLoS_formulation}, while the initial estimate of $\pp$ can be obtained from the \ac{AoD} estimates in \eqref{eq:LoS_formulation} or \eqref{eq:NLoS_formulation}.
This coarse position estimation problem can be solved by using geometric arguments by finding the least-squares intersection of the lines extending from the $\pprisr$ towards the \ac{UE} with estimated \ac{AoD} $\hat\thetab_r$, through
\begin{align}
    {\pp}_r = \pprisr + \beta_r \bu_r,~ r = 1, \cdots, R,
\end{align}
where $\bu_r= \bR_r^\trp\kk(\hat\thetab_r)/\norms{\bR_r^\trp\kk(\hat\thetab_r)}$ is the unitary direction vector and $\beta_r$ is unknown.

In order to locate the \ac{UE}, we need to find the closest point in the 3-D space to these lines. The least-square problem for the intersection of $R$ lines can be written as \cite{traa2013least}
\begin{align}
    &\hat\pp = \argmin_{\pp} \sum_{r=1}^R \norms{(\pp - \pprisr) - ((\pp - \pprisr)^\trp \bu_r)\bu_r}^2 \,.
\end{align}
It follows that \cite{alexandropoulos2022localization}
\begin{align} \label{eq:loc_gen}
    \hat\pp = &\Big(\sum_{r=1}^R \big(\eye -\bu_r\bu_r^\trp\big)\Big)^{-1} \Big(\sum_{r=1}^R \big(\eye -\bu_r\bu_r^\trp \big)\pprisr\Big). 
\end{align}


\subsection{Joint \ac{LoS} Detection and Localization} \label{sec:det}

In practice, the \ac{UE} may not know if the \ac{LoS} between itself and the \ac{BS} is blocked. To address this, we will introduce a \ac{GLRT}-based method to perform \ac{LoS} detection. Once the presence or absence of the \ac{LoS} path is known, we can choose the correct set of estimations. 
We formulate a hypothesis testing problem as follows:
\begin{align}\label{eq_hypotest-HW}
    \yy = \begin{cases}
	\yynlos(\zetab, \nu) + \nn,&~~ {\rm{under~\mathcal{H}_0}}  \\
	\yylos(\alphalos, \nu) + \yynlos(\zetab, \nu) + \nn ,&~~ {\rm{under~\mathcal{H}_1}} 
	\end{cases},
\end{align}
where  $\zetab = [\alphareo ~ \alphaimo ~ \cdots ~  \alphareR ~ \alphaimR ~ \thetab_1^\trp ~ \cdots ~ \thetabiR^\trp ]^\trp$. The null hypothesis ${\mathcal{H}_0} $ refers to the case where the \ac{LoS} path is blocked in \eqref{eq_yy}, while the alternate hypothesis 
 ${\mathcal{H}_1} $ refers to the case in which the \ac{LoS} path exists in \eqref{eq_yy}. In \eqref{eq_hypotest-HW},  
\begin{align}
   & \yylos(\alphalos, \nu) \triangleq \Psq  \alphalos  \,  \bb(\nu), \\
    &\yynlos(\zetab, \nu) \triangleq  \ \sum_{r=1}^{R} \Psq \alpharisr \, \xx_r(\thetabr) \odot \bb(\nu).
\end{align}
The \ac{GLRT} for the problem in \eqref{eq_hypotest-HW} can be expressed as \cite{van2004detection}
\begin{align}\label{eq_glrt}
   \llr(\yy) = \frac{ \max_{\alphalos, \zetab, \nu} p(\yy \, \lvert \, \mathcal{H}_1 ; \alphalos, \zetab, \nu ) }{\max_{\zetab, \nu} p(\yy \, \lvert \, \mathcal{H}_0 ; \zetab, \nu ) } \hdet \psi,
\end{align}
where $\psi$ is a threshold. 
Writing the log-likelihood ratio $\llrlog(\yy) \triangleq \sigma^2 \log \llr(\yy)$, we obtain
\begin{align}\label{eq:glrt2-HW}
    & \llrlog(\yy) = \left(\min_{\zetab, \nu} 
\norms{ \yy -  \yynlos(\zetab, \nu) }^2  -  \nonumber \right. \\ & \left.\min_{\alphalos, \zetab, \nu} 
\norms{ \yy - \yylos(\alphalos, \nu) - \yynlos(\zetab, \nu) }^2\right) \nonumber \\ & \hdet \psi^\prime =  \sigma^2 \log{\psi}.
\end{align}
There are two separate optimization problems to tackle in \eqref{eq:glrt2-HW}, identical to \eqref{eq:ML_LoS_RIS} and \eqref{eq:ML_NLoS_RIS}, which will  be solved in Sec.~\ref{sec:ChParamEst}: under \ac{LoS} in Sec.~\ref{subsec:LoS_localization} and NLoS in Sec.~\ref{subsec:NLoS_localization}. 
We can plug-in the resulting estimated \ac{LoS} and \ac{NLoS} parameter values into \eqref{eq:glrt2-HW} to perform \ac{LoS} detection. 
The algorithm is summarized in Algorithm~\ref{alg:det_GLRTinsp-HW}.
 \begin{algorithm}[t]
    \caption{Joint LoS Detection and Parameter Estimation Algorithm}\label{alg:det_GLRTinsp-HW}
    \hspace*{\algorithmicindent/2} \textbf{Input:} Received signal $\yy \in \mathbb{C}^{M}$ in \eqref{eq_yy}. \\
    \hspace*{\algorithmicindent/2} \textbf{Output:} Estimates $\hat{\nu}, \hat{\thetab}_1, \cdots, \hat{\thetab}_R$ 
    \begin{algorithmic}[1]
    \State Solve under~$\mathcal{H}_0$: Estimate $\hat{\nu}_{\rmnlos}, {\hat{\zetab}_\rmnlos}$ (see Section~\ref{subsec:NLoS_localization})
    \State Solve under~$\mathcal{H}_1$: 
    Estimate $\hat{\nu}_{\rmlos}, {\hat{\zetab}_\rmlos}$ and $\hat\alpha_0$ (see Section~\ref{subsec:LoS_localization})
    \State \ac{LoS} detection via \eqref{eq:glrt2-HW}: $ \llrlog(\yy) \hdet \sigma^2 \log{\psi}$.
    \State Based on the result, choose the final set of estimates accordingly.
    \end{algorithmic}
    \end{algorithm}

\section{Channel Parameter Estimation} \label{sec:ChParamEst}
In this section, we elaborate on the channel parameter estimation procedures that solve \eqref{eq:LoS_formulation} and \eqref{eq:NLoS_formulation}. First, we start under the assumption that \ac{LoS} exists in Sec.~\ref{subsec:LoS_localization}, tackling \eqref{eq:LoS_formulation}, then we continue with blocked \ac{LoS} assumption in Sec.~\ref{subsec:NLoS_localization}, focusing on \eqref{eq:NLoS_formulation}. The section concludes with a complexity analysis in Section~\ref{sec:complexity}.

\subsection{Channel Parameter Estimation under \ac{LoS}} \label{subsec:LoS_localization}
In this scenario, the \ac{LoS} path is the dominant path as the \ac{RIS}-induced paths are usually very weak. Hence, we treat the \ac{RIS} paths as noise for \ac{CFO} estimation, and then recover the signal per RIS, by harnessing the orthogonal RIS profiles. 
\subsubsection{CFO Estimation} \label{subsubsec:CFO_LoS}
 We form the \ac{ML} estimation of CFO with the assumption that the contribution from the \ac{RIS} paths is negligible. Under this assumption, 
we can rewrite the ML problem in \eqref{eq:LoS_formulation} as follows:
\begin{align}
    \hat \nu = \argmin_\nu\norms{\yy - \Psq \hat\alpha_0(\nu) \bb(\nu)}^2 = \argmax_\nu {\lvert\bb^\her(\nu)\yy \rvert^2},
\end{align}
which is a 1-D line search over the interval $-1/(2\Ts)<\nu <1/(2\Ts)$ to find coarse estimates. Then we can implement a 1-D quasi-Newton algorithm to find refined estimates of CFO.

\subsubsection{\ac{RIS} Separation}\label{subsub:RISSep}
Using the estimated \ac{CFO} $\hat\nu$, we wipe off its effect from the original observation in \eqref{eq_yy} as 
\begin{align} \label{eq:remove_CFO}
    \tilde{\yy} = \yy \odot \bb(-\hat \nu).
\end{align}
Then, assuming the residual CFO $\nu-\hat \nu$ is negligible, we separate each \ac{RIS} path by first forming the reshaped CFO-removed signal $\tilde{\bY}$ using \eqref{eq:YmatReshape} and then filtering it using \eqref{eq:time_coding}:
\begin{align} \label{eq:filterCFOremoved}
    \tilde\yy_{r} = \frac{1}{L} \tilde\bY^\trp \bc_r,~ r = 1,\cdots, R.
\end{align}
The CFO-removed, filtered contribution from the $\thn{r}$ \ac{RIS} can then be modeled explicitly as
\begin{align} \label{eq:RISith}
   \tilde \yy_r = \Psq \alpharisr \bar\bx_r(\thetabr) + \nn_r.
\end{align}
Here, $\nn_r$ contains noise plus any residual interference (due to residual CFO). The term $\bar\bx_r(\thetabr) = \bar \bW_r^\trp \aab(\thetabr) \in \complexsetone{M/L}$, and $\bar \bW_r \in \complexset{N}{M/L}$ contains the uncoded RIS phase profiles, i.e., 
\begin{align}
    &\bar \bW_r \triangleq [ \aab(\phibr) \odot \gammabrf{0} ~ \aab(\phibr) \odot \gammabrf{L} ~ \cdots ~ \aab(\phibr) \odot \gammabrf{M-L} ].
\end{align}


\subsubsection{\ac{AoD} Estimation per \ac{RIS}}
For \ac{CFO}-free $\thn{r}$ \ac{RIS} contribution in \eqref{eq:RISith}, we can formulate the corresponding \ac{ML} problem as
\begin{align}\label{eq:ML_LoS1}
    [\hat{\alpha}_r, \hat{\thetab}_r] = \argmin_{\alpharisr, \thetabr} \norms{\tilde\yy_r - \Psq \alpharisr  \bar\bx_r(\thetabr)}^2.
\end{align}
The path gain can be estimated in closed-form as a function of $\thetabr$ as follows:
\begin{align} \label{eq:alphaphat}
    \hat\alpha_r(\thetabr) = \frac{\bar\xx_r^\her(\thetabr) \tilde\yy_r}{ \Psq\norm{\bar\xx_r(\thetabr)}^2}.
\end{align}
Plugging \eqref{eq:alphaphat} into \eqref{eq:ML_LoS1} yields
\begin{align} \label{eq:ML_LoS2}
      \hat{\thetab}_r = & \argmin_{\thetabr} \norms{\tilde\yy_r -  \bar\bx_r(\thetabr) \frac{\bar\xx_r^\her(\thetabr) \tilde\yy_r}{ \norm{\bar\xx_r(\thetabr)}^2}}^2 \\
     = &  
     \argmin_{\thetabr} {\tilde\yy_r^\her\left(\eye -\frac{\bar\bx_r(\thetabr)\bar\xx_r^\her(\thetabr) }{\norm{\bar\xx_r(\thetabr)}^2}\right)\tilde\yy_r}.
\end{align}
Therefore, the \ac{ML} problem in \eqref{eq:ML_LoS1} reduces to
    \begin{align}\label{eq:ML_3}
     \hat{\thetab}_r = \argmax_{ \thetabr} \frac{|\tilde\yy_r^\her \bar\bx_r(\thetabr)|^2}{\norm{\bar\xx_r(\thetabr)}^2}.
\end{align}
For each \ac{RIS} $r$, this problem is solved separately by first performing a 2-D grid search for coarse estimation and then refining
by applying a 2-D quasi-Newton algorithm with the coarse estimate as the starting point.

The overall algorithm to solve \eqref{eq:LoS_formulation} is summarized in Algorithm~\ref{alg:LoS}. 
 \begin{algorithm}[t]
    \caption{\ac{LoS} Estimation Algorithm to Solve \eqref{eq:LoS_formulation}}\label{alg:LoS}
    \hspace*{\algorithmicindent/2} \textbf{Input:} Received signal $\yy \in \mathbb{C}^{M}$ in \eqref{eq_yy}.\\
    \hspace*{\algorithmicindent/2} \textbf{Output:} Estimates $\hat{\nu}, \hat{\thetab}_1, \cdots, \hat{\thetab}_R$. 
    \begin{algorithmic}[1]
    \State $\hat\nu = \argmax_{\nu} {\bb(\nu)^\her\yy}$.
   \State $\tilde{\yy} = \yy \odot \bb(-\hat \nu)$.
   \State $\tilde\bY = \text{reshape}(\tilde \yy, L, M/L)$ via \eqref{eq:YmatReshape}.
   \State \textbf{for} $r = 1, \cdots, R$ ~\textbf{do}
   \State \hspace{0.20cm} $\tilde\yy_{r} = {1}/{L} ~ \tilde\bY^\trp\bc_r $.
   \vspace{0.1cm} 
   \State \hspace{0.20cm}   
   $\hat{\thetab}_r = \argmax_{ \thetab} |\tilde\yy_r^\her \bar\bx_r(\thetab)|^2/\norm{\bar\xx_r(\thetab)}^2$ . 
   \State \textbf{end for}
  
    \end{algorithmic}
    \end{algorithm}


\subsection{Channel Parameter Estimation without \ac{LoS}} \label{subsec:NLoS_localization}
In \eqref{eq:NLoS_formulation}, the challenge is that unlike the previous scenario \eqref{eq:LoS_formulation}, in which there exists a dominant \ac{LoS} path, here the \ac{CFO} and \acp{AoD} are coupled in the received signal (as the CFO cannot be estimated using the \ac{LoS} path therefore its effect remains in the time-domain phase shifts). 
In this section, we present two low-complexity approaches to tackle this problem: the first one consists of $R$ 3-D searches, each including 1-D \ac{CFO} and 2-D \ac{RIS} \ac{AoD} search for each RIS. The second one involves a single 1-D \ac{CFO} estimation, followed by $R$ individual 2-D AoD estimations.

\subsubsection{ML Estimation} \label{subsec:NLOS_App1}

This approach is based on the observation that by conducting a 1-D search over the CFO, there will be an optimal value where all per-RIS observations can be decoupled. 
The approach operates as follows. For each trial value of the CFO $\nu \in \mathbb{C}_v$, we compute $\tilde{\yy}(\nu) = \yy \odot \bb(-\nu)$ as in \eqref{eq:remove_CFO}, where we make the dependence on $\nu$ explicit. Similarly, we compute  
$\tilde \yy_r(\nu )$ as in \eqref{eq:filterCFOremoved}. We then estimate the \ac{AoD} from each RIS 
$\hat{\thetab}_{1}(\nu ), \cdots, \hat{\thetab}_{R}(\nu )$ using \eqref{eq:ML_3}. Finally, we find the optimal $\nu $ (and \acp{AoD} as a result) from \eqref{eq:NLoS_formulation} as
\begin{align} \label{eq:NLoS_formulation-NU}
    &\hat\nu = \argmin_{\nu} \norms{\yy -\Psq\sum_{r=1}^{R}\hat\alpha_r( \nu)\xx_r(\hat{\thetab}_{r}(\nu))\odot\bb(\nu)}^2.
 \end{align}

The overall algorithm to solve \eqref{eq:NLoS_formulation} is summarized in Algorithm~\ref{alg:NLoS1}.

     \begin{algorithm}[t]
    \caption{NLoS Estimation Algorithm to Solve \eqref{eq:NLoS_formulation} - ML Estimation}\label{alg:NLoS1}
    \hspace*{\algorithmicindent/2} \textbf{Input:} Received signal $\yy \in \mathbb{C}^{M}$ in \eqref{eq_yy}.  \\
    \hspace*{\algorithmicindent/2} \textbf{Output:} Estimates $\hat{\nu}, \hat{\thetab}_1, \cdots, \hat{\thetab}_R$ 
    \begin{algorithmic}[1]
    \State \textbf{for} $\nu \in \mathbb{C}_v$ \textbf{do}
    \State \hspace{0.2cm} $\tilde{\yy}(\nu) = \yy \odot \bb(-\nu)$.
    \State \hspace{0.2cm} $\tilde\bY(\nu) = \text{reshape}(\tilde \yy(\nu) , L, M/L)$ via \eqref{eq:YmatReshape}.
    \State \hspace{0.2cm} \textbf{for} $r = 1, \cdots, R$ ~\textbf{do}
    \State \hspace{0.4cm} $\tilde\yy_{r}(\nu) = {1}/{L} ~\tilde\bY^\trp(\nu) \bc_r $. \vspace{0.1cm}
    \State \hspace{0.4cm} $\hat\thetab_{r}(\nu) = \argmax_{ \thetab} |\tilde\yy_r^\her(\nu) \bar\bx_r(\thetab)|^2/\norm{\bar\xx_r(\thetab)}^2$ .
    \State \hspace{0.2cm} \textbf{end for}
    \vspace{0.1cm}
    \State \textbf{end for}
    \State $\hat\nu = \argmin_{\nu \in \mathbb{C}_v} \norms{\yy -\Psq\sum_{r=1}^{R}\alpha_r(\nu)\xx_r(\hat\thetab_r(\nu))\odot\bb(\nu)}^2$.
    \end{algorithmic}
    \end{algorithm} 

\subsubsection{Low-complexity Unstructured Estimation} \label{subsec:NLOS_App2}

The ML estimation in \eqref{eq:NLoS_formulation-NU} requires a 3-D search and is thus computationally expensive. 
The motivation of this second approach is to de-couple the effect of \ac{CFO} and \acp{AoD} so as to estimate the \ac{CFO} with a 1-D search. To do so, we can express  into $\bY \in \complexset{L}{(M/L)}$ in \eqref{eq:YmatReshape} as
\begin{align} \label{alternate_model}
    \bY = \bD(\nu) \bC \bH(\Chib_\text{ch}) + \bN,
\end{align}
where $\Chib_\text{ch} = [\nu, \thetab_1^\trp, \cdots, \thetab_R^\trp]^\trp$, $\bD(\nu)=\text{diag}[1,e^{j2\pi \Ts \nu},\ldots,e^{j2\pi (L-1)\Ts \nu}]\in \complexset{L}{L}$ and
 the matrix $\bC \in \mathbb{C}^{L\times R}$ is the coding matrix which is composed of the coding vectors with length $L$ as below
\begin{align}
    \bC = [\bc_1, \cdots, \bc_R] \in \realset{L}{R} ~.
\end{align}
In addition, the matrix $\bH(\Chib_\text{ch}) \in \complexset{R}{(M/L)}$ is a function of both \ac{CFO} and \ac{AoD} with elements $h_{r,\ell} = [\bH(\Chib_\text{ch})]_{r,\ell}$, ($r = 1,\cdots,R, \ell = 1,\cdots,M/L$) as follows:
\begin{align}
   h_{r,\ell} = \Psq \alpha_{r} (\aab^\trp(\phib_{r})\odot{\bP_{r}}^\trp_{[:,\ell-1]})\aab(\thetab_{r}) e^{j 2 \pi (\ell-1) L \Ts \nu},
\end{align}
and $\bN \in \mathbb{C}^{L\times (M/L)}$ is the reshaped noise. With this re-modeling and ignoring the dependence of $\bH(\Chib_\text{ch})$ on the CFO, we de-couple the effect of \ac{CFO} and \acp{AoD} in the measurement model by factoring out matrix $\bD(\nu)$, which is a function of only CFO, and treating the matrix $\bH(\Chib_\text{ch})$ as an unstructured matrix. Therefore, it is enough to apply a 1-D grid search to estimate the \ac{CFO} through $\bD(\nu)$, without the need to estimate \acp{AoD} jointly. We will drop the dependence of $\bH$ on $\Chib_\text{ch}$ to simplify the notation.

To derive the \ac{CFO} estimator based on the model in \eqref{alternate_model}, we begin by vectorizing the observation as 
\begin{align}
    \yy & =\text{vec}(\bY) = (\eye_{M/L} \otimes (\bD(\nu)\bC))\bh + \bn = \bE(\nu)\bh + \bn,
\end{align}
where we used the property of the Kronecker product (eq. 520 in\cite{petersen2008matrix}). 
Here, $\bh =\text{vec}(\bH)\ \in \complexsetone{RM/L}$ and $\bE(\nu)=(\eye_{M/L} \otimes (\bD(\nu)\bC))\in \complexset{M}{(RM/L)}$. Accordingly, we formulate the estimator for jointly estimating $\nu$ and $\bh$ as 
\begin{align}
    [\hat\nu, \hat{\bh}] = \argmin_{\nu, \bh} \norm{\yy - \bE(\nu)\bh}^2,
\end{align}
where the conditional estimate of  $\bh$ is derived in closed form as
\begin{align}
    \hat{\bh}(\nu) & = ( \bE^\her(\nu)  \bE(\nu))^{-1} \bE^\her(\nu) \mathbf{y} = \frac{1}{L}\bE^\her(\nu) \mathbf{y},
\end{align}
since $ \bE^\her(\nu) \bE(\nu) = L\eye_{RM/L}$. 
Hence, we can write the estimate of $\nu $ as
\begin{align}
    \hat\nu & =  \argmin_{\nu} \norms{\yy - \frac{1}{L}\bE(\nu) \bE^\her(\nu) \yy}^2 \nonumber \\ 
    & = \argmin_{\nu} \norms{ \left( \eye_M - \frac{1}{L}\bE(\nu) \bE^\her(\nu)\right)\yy}^2 \nonumber \\ 
    & = \argmax_{\nu}\norms{\bE^\her(\nu)\yy}^2.
\end{align}
We can rewrite the objective function in matrix form using
\begin{align}
    \text{unvec}\left( \bE^\her(\nu)\yy \right)  = (\bD(\nu)\bC)^\her\bY = \bC^\her\bD^\her(\nu) \bY.
\end{align}
Here, the operator $\text{unvec}(\ba) = \bA$ transforms the vector $\ba \in \complexsetone{RM/L}$ to the matrix $\bA \in \complexset{R}{(M/L)}$. Hence, the optimization problem will be
\begin{align}
     \hat \nu = \argmax_{\nu}\norm{\bC^\her\bD^\her(\nu) \bY}^2_\mathrm{F},
\end{align}
where $\norm{\cdot}_\mathrm{F}$ stands for the Frobenius norm.

Note that the dependence of $\bH$ on \ac{CFO} and \acp{AoD} is ignored in the estimation algorithm. In other words, we do not leverage the full potential of the \ac{CFO}-dependent observations, which is the cost of detangling the effects of \ac{CFO} and \ac{AoD}. Therefore, this method is expected to be less accurate than the previous approach, but the complexity is lower and comparable to Algorithm~\ref{alg:LoS}. 

Once an estimate of the \ac{CFO} is obtained, it is possible to remove its effect from the observations as in \eqref{eq:remove_CFO}, and then apply the temporal decoding and $R$ 2-D grid searchs to find estimates of \acp{AoD}, as explained in Sec.~\ref{subsec:NLOS_App1}.
 The algorithm is summarized in Algorithm~\ref{alg:NLoS2}.

 \begin{remark}[Operation under the presence of LoS]
Note that \eqref{alternate_model} is a valid representation irrespective of whether the \ac{LoS} path exists or not. In case the \ac{LoS} exists, the coding matrix will change to $\bC = [\bc_0, \bc_1, \cdots, \bc_R]$ and $\bH \in \mathbb{C}^{ (R+1)\times M/L}$. Theoretically, it is possible to use the approach in Algorithm~\ref{alg:NLoS2} to estimate the \ac{CFO} in case \ac{LoS} exists, but under the assumption of dominant \ac{LoS}, the performance will be much lower than Algorithm~\ref{alg:LoS}.
\end{remark}

  \begin{algorithm}[t] 
    \caption{NLoS Estimation Algorithm to Solve \eqref{eq:NLoS_formulation} - Low-complexity Unstructured Estimation}\label{alg:NLoS2}
    \hspace*{\algorithmicindent/2} \textbf{Input:} Received signal $\yy \in \mathbb{C}^{M}$ in \eqref{eq_yy}. 
    \\
    \hspace*{\algorithmicindent/2} \textbf{Output:} Estimates $\hat{\nu}, \hat{\thetab}_1, \cdots, \hat{\thetab}_R$ 
    \begin{algorithmic}[1]
     \State $\bY = \text{reshape}( \yy , L, M/L)$ via \eqref{eq:YmatReshape}
     \State $\hat\nu =  \argmax_{\nu}\norm{\bC^\her\bD^\her(\nu) \bY}^2$
    \State $\tilde{\yy} = \yy \odot \bb(-\hat\nu)$
    \State $\tilde\bY(\nu) = \text{reshape}(\tilde \yy(\nu) , L, M/L)$ via \eqref{eq:YmatReshape}
   \State \textbf{for} $r = 1, \cdots, R$ ~\textbf{do}
   \State \hspace{0.20cm} $\tilde\yy_{r} = {1}/{L} ~\tilde\bY^\trp \bc_r$
   \State \hspace{0.2cm} 
   $\hat \thetab_r = \argmax_{ \thetab \in \thetab} |\tilde\yy_r^\her \bar\bx_r(\thetab)|^2/\norm{\bar\xx_r(\thetab)}^2$ 
   \State \textbf{end for}
    \end{algorithmic}
    \end{algorithm}
To address two NLoS estimation algorithms, we refer to Algorithm~\ref{alg:NLoS1} as NLoS-ML estimator and Algorithm~\ref{alg:NLoS2} as NLoS-LC (low complexity) estimator (with respect to NLoS-ML estimator).

\subsection{Complexity Analysis}\label{sec:complexity}

A brief complexity analysis is conducted. Assuming a fixed number of grid points $G$ in each dimension, the channel parameter estimation under \ac{LoS} (Algorithm~\ref{alg:LoS}) has a complexity of $\mathcal{O}(GM + R G^2 M + R^2)$, where the first term denotes the CFO estimation complexity, the second term corresponds the AoD estimation complexity and the third term shows the localization time based on \eqref{eq:loc_gen}. The NLoS-ML estimation approach 
has a complexity of $\mathcal{O}(R G^3 M + R^2)$ which reflects the fact that we need to perform 3-D estimations. The NLoS-LC 
has a complexity of $\mathcal{O}(GM + R G^2 M + R^2)$, which illustrates that the complexity returns to the level of under \ac{LoS} case with the expense of a reduction in performance which will be demonstrated in Sec.~\ref{sec:results}.


\section{Numerical Results}\label{sec:results}


In this section, we validate the proposed methods for \ac{LoS} detection, channel estimation and localization in both the presence and absence of the \ac{LoS} path between the \ac{BS} and the \ac{UE}. 
Several sensitivity studies will also be reported. 

\subsection{Scenario, Performance Metric and Simulation Setup}
We consider a scenario with  $R=2$ RIS, which is the minimal configuration needed to make the localization problem identifiable. The system parameters are summarized in Table~\ref{table:sim}. 
First, we will assume that we know whether the \ac{LoS} exists or not, therefore we will only focus on the performance of the estimators, with and without the presence of \ac{LoS}. Next, we will relax the aforementioned assumption to assess the performance of the detector together with the estimators. 

The path gains (due to propagation loss) $\alphalos$, $\alpha_1$ and $\alpha_2$ in \eqref{eq_yy} are determined based on the \ac{FSPL} model, containing random phases between $[0, 2\pi)$, similar to \cite{ris_propagation_modeling_2020}:
\begin{align}
    |{\alpha_0}| & = \frac{\lambda}{4\pi d_\text{BS-UE}}  \\
     |{\alpha_r}| & = \frac{\lambda^2}{16\pi^2 {d_\text{BS-RISr}}d_\text{RISr-UE}}, r = 1,2,
\end{align}
where the effective area of each RIS element has been assumed to be $\lambda^2/4\pi$, $d_\text{I-J}$, I = \{BS, RISr\}, J = \{UE, RISr\} is the distance between the entity I and the entity J.
In order to assess the performance of the estimators, we calculate the \ac{RMSE} of our estimates averaged over $100$ Monte-Carlo trials, and then compare them with \ac{CRB} as the benchmark. More details on how we derive CRB of different unknowns in our setup are provided in Appendix~\ref{app:FIM}. To assess the performance of the detector, the false alarm probability is calculated over 500 Monte-Carlo trials, and the threshold is chosen numerically to attain a detection probability close to 1. The detection performance turns out to be relatively insensitive to the value of the threshold for the power levels considered in the simulations. 

\begin{table}
\renewcommand{\arraystretch}{1.3}
\caption{Simulations Parameters}
\label{table:sim}
\centering
\begin{tabular}{l c c}
\hline
\bfseries Parameters & \bfseries Symbol & \bfseries Value \\
\hline\hline
Number of RISs & $R$ & 2\\
Wavelength & $\lambda$ & 1 cm\\
Sampling time & $\Ts$ & 10 $\mu$sec\\
\ac{RIS} dimensions & $N$ & $64\times 64$ \\
RIS element spacing & $d$ & 0.5 cm\\
Speed of Light & $c$ & $3 \times 10^8$ m/s\\
Noise PSD & $N_0$ & -174 dBm/Hz \\
UE's Noise Figure & $n_f$ & 8 dB \\
Noise power & $\sigma^2$ & $ N_0 /\Ts \times n_f$ W\\
Number of Transmissions & $M$ & 256 \\
BS position & $\ppbs$ & [0, 0, 0] m \\
UE position & $\pp$ & [5, 2, 0.5] m \\
RIS1 position & $\ppriso$ & [10, -10, 0] m \\
RIS2 position & $\pprist$ & [0, 10, 0] m \\
RIS1 rotation matrix & $\bR_1$ & $\bR_z(\theta = 0)$\\
RIS2 rotation matrix & $\bR_2$ & $\bR_z(\theta = \pi)$ \\
\hline
\end{tabular}
\end{table} 



\subsection{Localization under \ac{LoS}}\label{sec_results_los}
In this section, we investigate the accuracy of our proposed estimation algorithm, Algorithm~\ref{alg:LoS}, in Sec.~\ref{subsec:LoS_localization} under the assumption that a dominant \ac{LoS} path exists between the \ac{BS} and the \ac{UE}. To do so, we find the \ac{RMSE} of the \ac{CFO}, \acp{AoD}, and position, and compare them to the \acp{CRB}. For achieving higher precision estimates, we employ the \textit{quasi-Newton} algorithm to perform 1-D CFO refinement and two 2-D AoD refinements.
\subsubsection{Channel Parameter Estimation}

\begin{figure}
\centering



\begin{tikzpicture}
[scale=1\columnwidth/10cm,font=\footnotesize]
\begin{axis}[
    width=10cm,
    height=7cm,
    ymode=log,
    log ticks with fixed point,
    xmin=0,
    xmax=40,
    ymin=0.001,
    ymax=0.1,
    xlabel={$P$ [dBm]},
    ylabel={RMSE/CRB [Hz]},  
    grid=major,
    legend style={at={(1,1)},anchor=north east, scale = 1,  fill opacity=0.5,
            draw opacity=1,   
            text opacity=1},
    legend cell align=left
]
\addplot[draw=red,mark=o, line width=1pt, mark repeat=2] table[x=P, y=CFO_ref, col sep=comma]{tikzfig/data/RMSE_LoS_vs_PWR.txt};
\addplot[draw=black, dashed, line width=1pt, mark repeat=2] table[x=P, y=CFO, col sep=comma]{tikzfig/data/CRB_LoS_vs_PWR.txt};
\addlegendentry{RMSE}
\addlegendentry{CRB}
\end{axis}
\end{tikzpicture}

\vspace{-4mm}
\caption{LoS case, CFO estimation performance vs. transmit power.}
\label{fig:CFO_LoS_vs_PWR}
\end{figure}
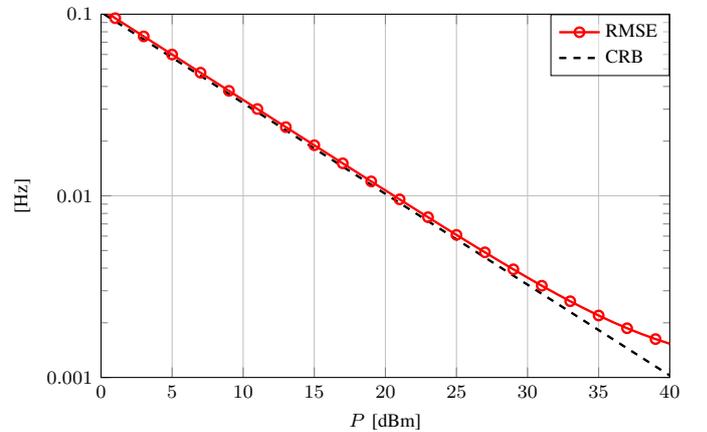

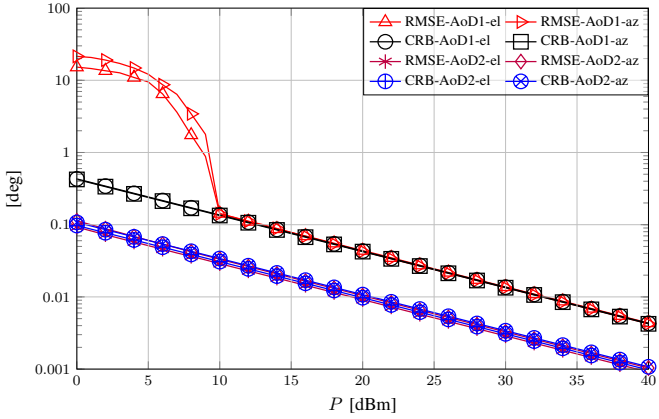
\begin{figure}
\centering
\pgfkeys{/pgf/number format/.cd,fixed,precision=3}
 \usepgfplotslibrary{groupplots}
\pgfplotsset{
    width=11cm,
    height=7cm,
    compat=1.3, 
   every axis/.append style={scale only axis, axis on top,
    xmin=0, xmax=40, ymin=0.001, ymax=100,     
    ymode=log,
    log ticks with fixed point, 
    ylabel={RMSE/CRB [deg]},  
    xlabel={$P$ [dBm]},
    xlabel style={font=\normalsize},
    ylabel style={font=\normalsize},
    grid=major,
     legend style={at={(1,1)},anchor=north east, scale = 1.1,  fill opacity=0.5,
            draw opacity=1,   
            text opacity=1, legend columns = 2},
     legend cell align=left
    }
}
\begin{tikzpicture}
[scale=1\columnwidth/13cm,font=\footnotesize]
    \begin{axis}
        \addplot[line width=0.7pt][draw=red, mark=triangle, mark repeat=2, mark options={scale=2}] table[x=P, y=AoD1_1_ref, col sep=comma]{tikzfig/data/RMSE_LoS_vs_PWR.txt}; 
        \addplot[line width=0.7pt, mark repeat=2][draw=red, mark=triangle, mark options={rotate=270, scale = 2}] table[x=P, y=AoD1_2_ref, col sep=comma]{tikzfig/data/RMSE_LoS_vs_PWR.txt};
        
        \addplot[line width=0.7pt][draw=black, mark = o, mark repeat=2, mark options={scale=2}] table[x=P, y=AoD1_1, col sep=comma]{tikzfig/data/CRB_LoS_vs_PWR.txt};
        \addplot[line width=0.7pt][draw=black, mark = square, mark repeat=2, mark options={scale=2}] table[x=P, y=AoD1_2, col sep=comma]{tikzfig/data/CRB_LoS_vs_PWR.txt};
        
        \addplot[line width=0.7pt][draw=purple, mark=asterisk, mark repeat=2, mark options={scale=2}] table[x=P, y=AoD2_1_ref, col sep=comma]{tikzfig/data/RMSE_LoS_vs_PWR.txt};
        \addplot[line width=0.7pt][draw=purple, mark=diamond, mark repeat=2, mark options={scale=2}] table[x=P, y=AoD2_2_ref, col sep=comma]{tikzfig/data/RMSE_LoS_vs_PWR.txt};
             
        \addplot[line width=0.7pt][draw=blue, mark = oplus, mark repeat=2, mark options={scale=2}] table[x=P, y=AoD2_1, col sep=comma]{tikzfig/data/CRB_LoS_vs_PWR.txt};
        \addplot[line width=0.7pt][draw=blue, mark = otimes, mark repeat=2, mark options={scale=2}] table[x=P, y=AoD2_2, col sep=comma]{tikzfig/data/CRB_LoS_vs_PWR.txt};
        \addlegendentry{RMSE-AoD1-el}
        \addlegendentry{RMSE-AoD1-az}
        \addlegendentry{CRB-AoD1-el}     
        \addlegendentry{CRB-AoD1-az}
        
        \addlegendentry{RMSE-AoD2-el}
        \addlegendentry{RMSE-AoD2-az}
        \addlegendentry{CRB-AoD2-el}     
        \addlegendentry{CRB-AoD2-az}        
    \end{axis}

\end{tikzpicture}\vspace{-4mm}
\caption{LoS case, AoD1 and AoD2 estimation performance vs. transmit power.}
\label{fig:AoD_LoS_vs_PWR}
\end{figure}

In Fig.~\ref{fig:CFO_LoS_vs_PWR}, the RMSE of CFO together with the CRB are depicted versus the transmit power $P$ . The \ac{CFO} is fixed to $-40$ kHz\footnote{Later, in Fig. \ref{fig:CFO_sens}, we show that the CRB and RMSE from our algorithms remain nearly constant across different CFO values, making the choice of $\nu = -40$ kHz arbitrary.}. 
Since the \ac{LoS} is dominant, even at low transmit power we manage to 
achieve the bound. 
Fig.~\ref{fig:AoD_LoS_vs_PWR} shows the RMSE of \acp{AoD} versus transmit power for RIS1 and RIS2. We will refer to \ac{AoD} for RIS1 as AoD1 and AoD for RIS2 as AoD2. 
It can be observed that the bounds related to the azimuth and elevation angles of AoD2 are smaller than those of AoD1, and with our algorithm it is possible to achieve the CRB at lower transmit power for AoD2 than for AoD1. The reason is as follows: According to Table~\ref{table:sim}, RIS2 is closer to the BS and the UE than RIS1, suggesting that $\alpha_1 < \alpha_2$. Since the received power at the UE through BS-RIS2-UE path is larger than through BS-RIS1-UE, it is easier to estimate AoD from RIS2 than from RIS1, resulting in lower CRB and RMSE. Overall, Fig.~\ref{fig:CFO_LoS_vs_PWR} and Fig.~\ref{fig:AoD_LoS_vs_PWR} demonstrate the effectiveness of the channel estimation algorithm in Algorithm~\ref{alg:LoS}, indicating convergence to the theoretical bounds already at low transmit powers.

\subsubsection{Localization}
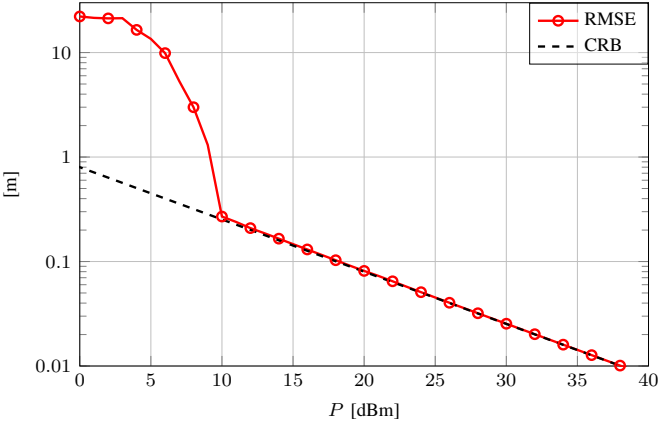
\begin{figure}
\centering


\begin{tikzpicture}
[scale=1\columnwidth/10cm,font=\footnotesize]
set(gca,'fontsize',24)
\begin{axis}[
    width=10cm,
    height=7cm,
    ymode=log,
    log ticks with fixed point,
    xmin=0,
    xmax=40,
    ymin=0.01,
    ymax=30,
    xlabel={$P$ [dBm]},
    ylabel={RMSE/CRB [m]},  
    grid=major,
    legend style={at={(1,1)},anchor=north east, fill opacity=0.5,
            draw opacity=1,   
            text opacity=1, scale = 1},
    legend cell align=left
]
\addplot[draw=red, line width=1pt,mark=o, mark repeat=2] table[x=P, y=pos, col sep=comma]{tikzfig/data/RMSE_LoS_vs_PWR.txt};
\addplot[draw=black, line width=1pt, mark repeat=2, dashed] table[x=P, y=PEB, col sep=comma]{tikzfig/data/CRB_LoS_vs_PWR.txt};
\addlegendentry{RMSE}
\addlegendentry{CRB}
\end{axis}
\end{tikzpicture}

\vspace{-4mm}
\caption{LoS case, position estimation performance vs. transmit power.}
\label{fig:pos_LoS_vs_PWR}
\end{figure}

Fig.~\ref{fig:pos_LoS_vs_PWR} shows the RMSE and the CRB on location estimation against the transmit power. It can be observed that the proposed localization algorithm in Sec.~\ref{sec_loc} can attain the bound at the transmit power for which the RMSE of both \acp{AoD} have attained their respective bounds, as expected. 

\vspace{0.1in}

\subsection{Localization without LoS}
In this subsection, we perform the same experiments as explained in the previous subsection for measurements that do not contain the \ac{LoS} path contribution in \eqref{eq_yy}, using the two approaches, Algorithm~\ref{alg:NLoS1} and Algorithm~\ref{alg:NLoS2}, as explained in Sec.~\ref{subsec:NLoS_localization} and compare the results. To obtain more accurate estimates, we utilize the \textit{quasi-Newton} algorithm to perform 5-D CFO/AoD refinement using the objective function in \eqref{eq:NLoS_formulation}.

\subsubsection{Channel Parameter Estimation}
\begin{figure}
\centering
\begin{tikzpicture}
[scale=1\columnwidth/10cm,font=\footnotesize]
\begin{axis}[
    width=10cm,
    height=7cm,
    ymode=log,
    log ticks with fixed point,
    xmin= 0,
    xmax=40,
    ymin=0.1,
    ymax=20000,
    xlabel={$P$ [dBm]},
    ylabel={RMSE/CRB [Hz]},   
    grid=major,
    legend style={at={(1,1)},anchor=north east, scale = 1, fill opacity=0.5,
            draw opacity=1,   
            text opacity=1},
    legend cell align=left
]
\addplot[draw=green!50!black, dashed, mark=oplus, line width=1pt, mark repeat=2, mark options=solid] table[x=P, y=CFO_ref, col sep=comma]{tikzfig/data/RMSE_nLoS_app1_vs_PWR.txt};
\addplot[draw=orange, mark=square, line width=1pt, mark repeat=2] table[x=P, y=CFO_ref, col sep=comma]{tikzfig/data/RMSE_nLoS_app2_vs_PWR.txt};
\addplot[draw=blue, dashed, line width=1pt, mark repeat=2] table[x=P, y=CFO, col sep=comma]{tikzfig/data/CRB_nLoS_app1_vs_PWR.txt};
\addlegendentry{RMSE-ML}
\addlegendentry{RMSE-LC}
\addlegendentry{CRB}
\end{axis}
\end{tikzpicture}

\vspace{-4mm}
\caption{NLoS case, CFO estimation performance vs. transmit power.}
\label{fig:CFO_nLoS_vs_PWR}
\end{figure}
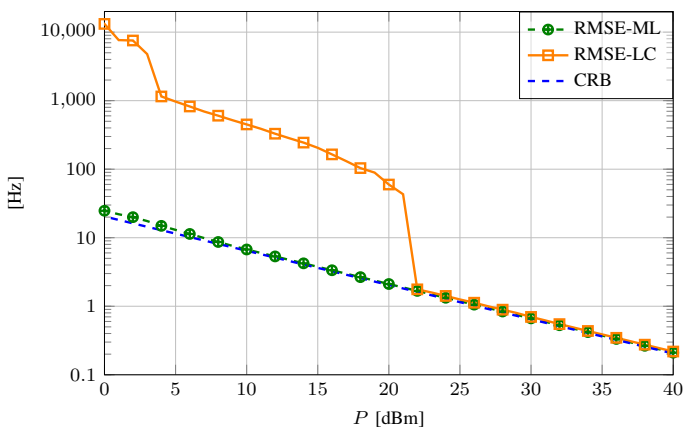

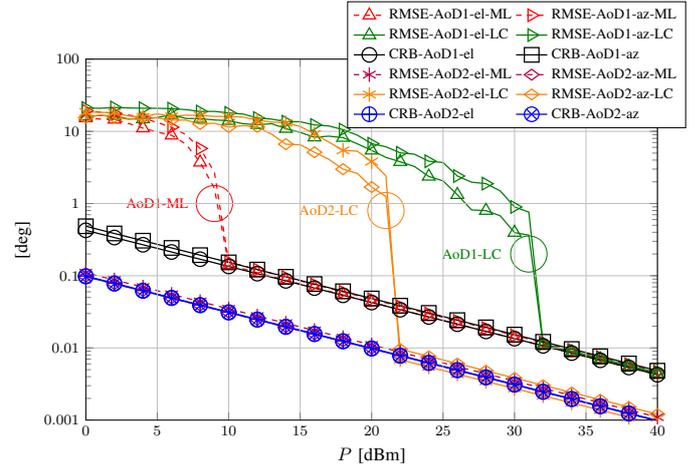
\begin{figure}
\centering
\pgfkeys{/pgf/number format/.cd,fixed,precision=3}
 \usepgfplotslibrary{groupplots}
\pgfplotsset{
    width=11cm,
    height=7cm,
    compat=1.3, 
   every axis/.append style={scale only axis, axis on top,
    xmin=0, xmax=40, ymin=0.001, ymax=100,     
    ymode=log,
    log ticks with fixed point, 
    ylabel={RMSE/CRB [deg]},  
    xlabel={$P$ [dBm]},
    xlabel style={font=\normalsize},
    ylabel style={font=\normalsize},
    grid=major,
     legend style={at={(1.05,1.16)},anchor=north east, scale = 0.95, legend columns = 2},
     legend cell align=left
    }
}
\begin{tikzpicture}
[scale=1\columnwidth/13cm,font=\footnotesize]
    \begin{axis}
         
        \addplot[draw=red, mark = triangle, dashed, line width=0.7pt, mark repeat=2, mark options={scale = 2, solid}] table[x=P, y=AoD1_1_ref, col sep=comma]{tikzfig/data/RMSE_nLoS_app1_vs_PWR.txt};

        \addplot[draw=red, mark = triangle, dashed, line width=0.7pt, mark options={rotate=270, scale = 2, solid}, mark repeat=2] table[x=P, y=AoD1_2_ref, col sep=comma]{tikzfig/data/RMSE_nLoS_app1_vs_PWR.txt};
              
        \addplot[draw=green!50!black, mark = triangle, line width=0.7pt, mark repeat=2, mark options={scale = 2}] table[x=P, y=AoD1_1_ref, col sep=comma]{tikzfig/data/RMSE_nLoS_app2_vs_PWR.txt};
        
        \addplot[draw=green!50!black, mark = triangle, mark options={rotate=270, scale = 2}, line width=0.7pt, mark repeat=2] table[x=P, y=AoD1_2_ref, col sep=comma]{tikzfig/data/RMSE_nLoS_app2_vs_PWR.txt};
    
        \addplot[draw=black, mark = o, line width=0.7pt, mark repeat=2, mark options={scale = 2}] table[x=P, y=AoD1_1, col sep=comma]{tikzfig/data/CRB_nLoS_app1_vs_PWR.txt};  

        \addplot[draw=black, mark = square, line width=0.7pt, mark repeat=2, mark options={scale = 2}] table[x=P, y=AoD1_2, col sep=comma]{tikzfig/data/CRB_nLoS_app1_vs_PWR.txt};
         
         \addplot[draw=purple, dashed, mark = asterisk, line width=0.7pt, mark repeat=2, mark options={scale = 2, solid}] table[x=P, y=AoD2_1_ref, col sep=comma]{tikzfig/data/RMSE_nLoS_app1_vs_PWR.txt};

        \addplot[draw=purple, mark = diamond, dashed, mark options={rotate=270, scale = 2, solid}, line width=0.7pt, mark repeat=2] table[x=P, y=AoD2_2_ref, col sep=comma]{tikzfig/data/RMSE_nLoS_app1_vs_PWR.txt};
              
        \addplot[draw=orange, mark = asterisk, line width=0.7pt, mark repeat=2, mark options={scale = 2}] table[x=P, y=AoD2_1_ref, col sep=comma]{tikzfig/data/RMSE_nLoS_app2_vs_PWR.txt};
        
        \addplot[draw=orange, mark = diamond, mark options={rotate=270, scale = 2}, line width=0.7pt, mark repeat=2] table[x=P, y=AoD2_2_ref, col sep=comma]{tikzfig/data/RMSE_nLoS_app2_vs_PWR.txt};
    
        \addplot[draw=blue, mark = oplus, line width=0.7pt, mark repeat=2, mark options={scale = 2}] table[x=P, y=AoD2_1, col sep=comma]{tikzfig/data/CRB_nLoS_app1_vs_PWR.txt};  

        \addplot[draw=blue, mark = otimes, line width=0.7pt, mark repeat=2, mark options={scale = 2}] table[x=P, y=AoD2_2, col sep=comma]{tikzfig/data/CRB_nLoS_app1_vs_PWR.txt};

        \node[circle, draw = red, minimum size=0.7cm] at (axis cs:9,1) {};

        \node at (axis cs:5,1) [color=red]{AoD1-ML};

        \node[circle, draw = orange, minimum size=0.7cm] at (axis cs:21,0.8) {};

        \node at (axis cs:17,0.8) [color=orange]{AoD2-LC};

        \node[circle, draw = green!50!black, minimum size=0.7cm] at (axis cs:31,0.2) {};

        \node at (axis cs:27,0.2) [color=green!50!black]{AoD1-LC};

        \addlegendentry{RMSE-AoD1-el-ML}
        \addlegendentry{RMSE-AoD1-az-ML}
        \addlegendentry{RMSE-AoD1-el-LC}
        \addlegendentry{RMSE-AoD1-az-LC}
        \addlegendentry{CRB-AoD1-el}     
        \addlegendentry{CRB-AoD1-az}    

        \addlegendentry{RMSE-AoD2-el-ML}
        \addlegendentry{RMSE-AoD2-az-ML}
        \addlegendentry{RMSE-AoD2-el-LC}
        \addlegendentry{RMSE-AoD2-az-LC}
        \addlegendentry{CRB-AoD2-el}     
        \addlegendentry{CRB-AoD2-az}
    \end{axis}

\end{tikzpicture}
\caption{NLoS case, AoD1 and AoD2 estimation performance vs. transmit power.}
\label{fig:AoD_nLoS_vs_PWR}
\end{figure}

Fig.~\ref{fig:CFO_nLoS_vs_PWR} and~\ref{fig:AoD_nLoS_vs_PWR} show the RMSE of \ac{CFO} and \acp{AoD} versus transmit power according to the NLoS-ML estimator (Algorithm~\ref{alg:NLoS1}) and low-complexity unstructured (marked by NLoS-LC - Algorithm~\ref{alg:NLoS2}) estimator. 
The results show that with NLoS-ML estimator, it is possible to touch the bound at lower transmit power comparing to the NLoS-LC estimator. This advantage is achieved at the expense of higher complexity comparing to the NLoS-LC estimator.

\vspace{4mm}
\subsubsection{Localization}
Accordingly, Fig.~\ref{fig:pos_nLoS_vs_PWR} represents the positional \ac{RMSE} versus transmit power for two approaches. It can be observed that it is possible to achieve the bound at lower transmit power in NLoS-ML estimator than in NLoS-LC estimator, as previous results suggested. Similar to the results presented in Sec.~\ref{sec_results_los} for the scenario with \ac{LoS}, Fig.~\ref{fig:CFO_nLoS_vs_PWR}, Fig.~\ref{fig:AoD_nLoS_vs_PWR} and Fig.~\ref{fig:pos_nLoS_vs_PWR} reveal the effectiveness of Algorithm~\ref{alg:NLoS1} and Algorithm~\ref{alg:NLoS2}, as well as the localization algorithm in Sec.~\ref{sec_loc} to solve \eqref{eq:opt_localizationnLoS}.
\begin{remark}
The sharp decline in RMSE observed in figures such as Fig.~\ref{fig:AoD_LoS_vs_PWR}, Fig.~\ref{fig:pos_LoS_vs_PWR}, etc. is attributed to a well-known phenomenon known as waterfall behavior \cite{gaudio2020joint, rodriguez2021joint}. This phenomenon occurs when increasing the transmit power leads to a point where the signal begins to dominate the noise. Consequently, the RMSE converges to the CRB, resulting in a rapid drop-off in RMSE \cite{Kamran_JSTSP_SISO_RIS, wymeersch2020radio, tdm_mimo_radar_JSTSP_2021}.
\end{remark}
\begin{remark}
  It is worth noting that while the presented optimization problems in Sec. \ref{sec:highLevel} are not globally convex, the objective functions decrease smoothly toward the ML solution in the region of interest. This structure allows gradient descent to reliably converge to the correct solution when initialized close to the optimum. The effectiveness of our initialization methods is confirmed by the fact that the final estimates consistently achieve the CRB.
\end{remark}
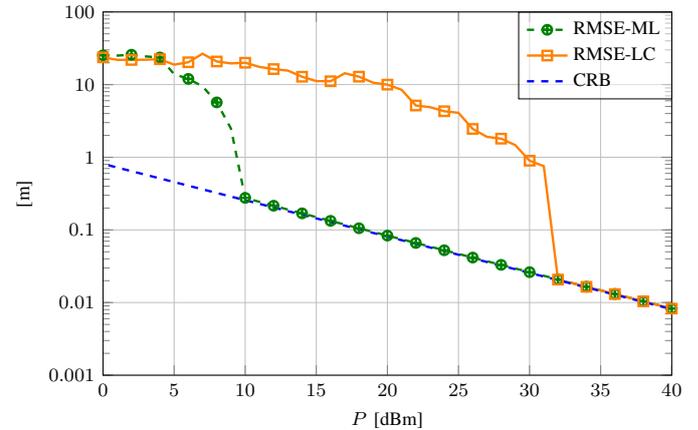
\begin{figure}
\centering


\begin{tikzpicture}
[scale=1\columnwidth/10cm,font=\footnotesize]
set(gca,'fontsize',24)
\begin{axis}[
    width = 10cm,
    height = 7cm,
    ymode=log,
    log ticks with fixed point,
    xmin=0,
    xmax=40,
    ymin=0.001,
    ymax=100,
    xlabel={$P$ [dBm]},
    ylabel={RMSE/CRB [m]},  
    grid=major,
    legend style={nodes={scale=1, transform shape}, at={(0.73,0.88)},anchor=west, fill opacity=0.5,
            draw opacity=1,   
            text opacity=1},
    legend cell align=left
]
\addplot[draw=green!50!black, mark=oplus, dashed, line width=1pt, mark repeat=2, mark options=solid] table[x=P, y=pos, col sep=comma]{tikzfig/data/RMSE_nLoS_app1_vs_PWR.txt};
\addplot[draw=orange, mark=square, line width=1pt, mark repeat=2] table[x=P, y=pos, col sep=comma]{tikzfig/data/RMSE_nLoS_app2_vs_PWR.txt};
\addplot[draw=blue, line width=1pt, dashed, mark repeat=2] table[x=P, y=PEB, col sep=comma]{tikzfig/data/CRB_nLoS_app1_vs_PWR.txt};
\addlegendentry{RMSE-ML}
\addlegendentry{RMSE-LC}
\addlegendentry{CRB}
\end{axis}
\end{tikzpicture}

\caption{NLoS case, position estimation performance vs. transmit power.}
\label{fig:pos_nLoS_vs_PWR}
\end{figure}

\subsection{Joint \ac{LoS} Detection and localization}
Here, we analyze the results of the presented \ac{LoS} detector (Algorithm~\ref{alg:det_GLRTinsp-HW}) using two alternative NLoS estimators, i.e. NLoS-ML estimator (Algorithm~\ref{alg:NLoS1}) and NLoS-LC estimator (Algorithm~\ref{alg:NLoS2}).
  
\begin{figure}
\centering


\begin{tikzpicture}
[scale=1\columnwidth/10cm,font=\footnotesize]
\begin{axis}[
    xmin=0,
    xmax=40,
    ymin=0,
    width = 10cm,
    height = 7cm,
    ymax=0.4,
    xlabel={$P$ [dBm]},
    ylabel={False alarm probability},  
    grid=major,
    legend style={at={(1,1)},anchor=north east, scale = 1, fill opacity=0.5,
            draw opacity=1,   
            text opacity=1},
    legend cell align=left
]
\addplot[draw=green!50!black, mark=oplus, dashed, line width=1pt, mark repeat=2, mark options=solid] table[x=P, y=p_fa_sim_GLRTinspired_app1_nLoS, col sep=comma]{tikzfig/data/pfa.txt};
\addplot[draw=orange, mark = square, line width=1pt, mark repeat=2] table[x=P, y=p_fa_sim_GLRTinspired_app2_NLoS, col sep=comma]{tikzfig/data/pfa.txt};
\addlegendentry{ML}
\addlegendentry{LC}
\end{axis}
\end{tikzpicture}

\vspace{-4mm}
\caption{False alarm probability as a function of the transmit power. }
\label{fig:pfa}
\end{figure}
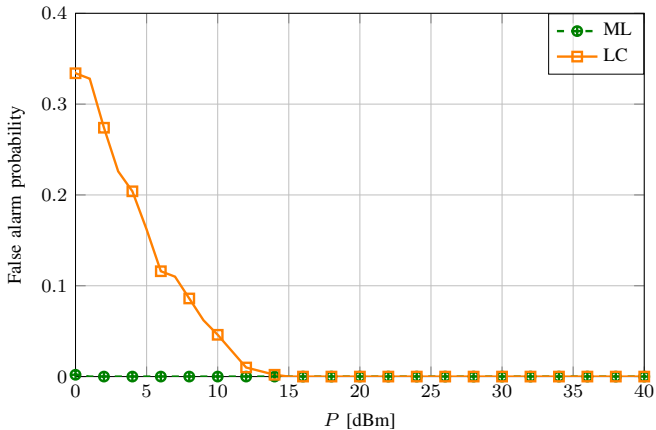

    Fig.~\ref{fig:pfa} shows the false alarm probability vs. transmit power for different estimators presented in Sec.~\ref{subsec:NLoS_localization}. By false alarm, we refer to the case that a \ac{LoS} path does not exist but the detector declares that a \ac{LoS} path exists. While using NLoS-ML estimator (Algorithm~\ref{alg:NLoS1}) in case of NLoS hypothesis yields almost all zero false alarm probability within the desired transmit power range, using low-complexity approach (Algorithm~\ref{alg:NLoS2}) results in non-zero false alarm probability. This outcome is due to the fact that the quality of our detector is closely tied to the accuracy of the corresponding estimator, which is noticeably compromised in case of low-complexity estimator. 

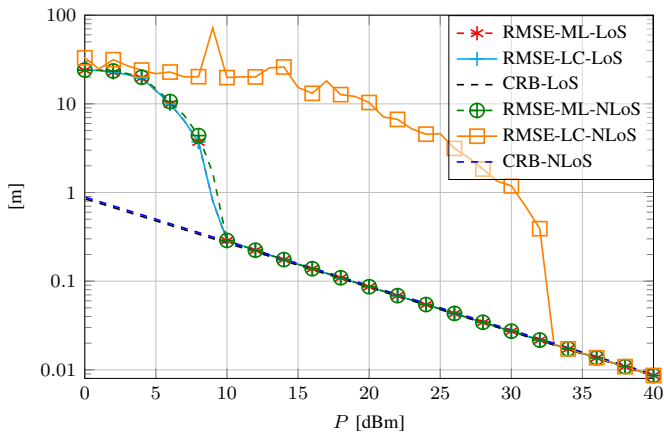
\begin{figure}
\centering


\begin{tikzpicture}
[scale=1\columnwidth/10cm,font=\footnotesize]
\begin{axis}[
    width = 10cm,
    height = 7cm,
    ymode=log,
    log ticks with fixed point,
    xmin=0,
    xmax=40,
    ymin=0.008,
    ymax=100,
    xlabel={$P$ [dBm]},
    ylabel={RMSE/CRB [m]},  
    grid=major,
    legend style={at={(1,1)},anchor=north east, scale = 1,  fill opacity=0.5,
            draw opacity=1,   
            text opacity=1,
    legend cell align=left}
]
\addplot[draw=red, mark = asterisk, dashed, line width=0.7pt, mark repeat=2, mark options={scale = 1.5, solid}] table[x=P, y=pos, col sep=comma]{tikzfig/data/RMSE_det_LoS_GLRTinspired_App1.txt};
\addplot[draw=cyan, mark = +, line width=0.7pt, mark repeat=2, mark options={scale = 1.5}] table[x=P, y=pos, col sep=comma]{tikzfig/data/RMSE_det_LoS_GLRTinspired_App2.txt};
\addplot[draw=black, dashed, line width=0.7pt, mark repeat=2] table[x=P, y=PEB, col sep=comma]{tikzfig/data/CRB_det_LoS.txt};
\addplot[draw=green!50!black, mark = oplus, dashed, line width=0.7pt, mark repeat=2, mark options={scale = 1.5, solid}] table[x=P, y=pos, col sep=comma]{tikzfig/data/RMSE_det_noLoS_GLRTinspired_App1.txt};
\addplot[draw=orange, mark = square, line width=0.7pt, mark repeat=2, mark options={scale = 1.5}] table[x=P, y=pos, col sep=comma]{tikzfig/data/RMSE_det_noLoS_GLRTinspired_App2.txt};
\addplot[draw=blue, line width=0.7pt, mark repeat=2, dashed] table[x=P, y=PEB, col sep=comma]{tikzfig/data/CRB_det_noLoS.txt};
\addlegendentry{RMSE-ML-LoS}
\addlegendentry{RMSE-LC-LoS}
\addlegendentry{CRB-LoS}
\addlegendentry{RMSE-ML-NLoS}
\addlegendentry{RMSE-LC-NLoS}
\addlegendentry{CRB-NLoS}
\end{axis}
\end{tikzpicture}

\vspace{-4mm}
\caption{Position estimation performance vs. transmit power, using LoS detector, in LoS and NLoS cases.}
\label{fig:pos_det_LoS_nLoS}
\end{figure}

    Fig.~\ref{fig:pos_det_LoS_nLoS} presents the positioning RMSE and CRB versus transmit power 
    with \ac{LoS} and without \ac{LoS} respectively. It should be pointed out that the CRB in LoS and NLoS scenarios are almost the same, reflecting the fact that the \ac{LoS} path does not convey any significant localization information. When \ac{LoS} exists, the performance of the detector using either estimator are similar, since the detector is able to detect the presence of \ac{LoS}, then the positional RMSE is based on the output of the LoS algorithm and is similar to Fig.~\ref{fig:pos_LoS_vs_PWR}. On the other hand, when LoS does not exist, by using NLoS-ML estimator the detector is able to verify that LoS path does not exist even at low transmit power (cf. Fig.~\ref{fig:pfa}), therefore the positional RMSE is calculated using the output of the NLoS algorithm and the results are similar to Fig.~\ref{fig:pos_nLoS_vs_PWR} with NLoS-NLoS-ML estimator. In case of using NLoS-LC estimator, we observe non-zero false alarm probability in Fig.~\ref{fig:pfa}, but according to Fig.~\ref{fig:pos_nLoS_vs_PWR}, using NLoS-LC estimator already results in poor RMSE at medium transmit power, therefore the non-zero false alarm probability does not considerably affect the estimation accuracy comparing to Fig.~\ref{fig:pos_nLoS_vs_PWR}.


\subsection{Comparison with Existing Methods}
Here, we compare our proposed CFO-aware localization method to a CFO-agnostic localization method (similar to the one proposed in \cite{keykhosravi2021siso}) to highlight the effect of CFO estimation on the quality of our positioning outcome. 
The result is shown in Fig.~\ref{fig:CFO_sens} for both the LoS scenario (denoted by the legend 'RMSE-LoS-noCFO') and NLoS scenario (denoted by the legend 'RMSE-NLoS-noCFO'). The transmit power is fixed to $P = 35$ dBm. It can be observed that when we ignore the presence of CFO and bypass the CFO estimation step, the RMSE of the estimated position is close to the bound only for very small CFO and deviates from the CRB for larger \ac{CFO}, with larger deviation when \ac{LoS} path exists. The reason is that the \ac{LoS} component behaves as a strong source of interference when the \ac{CFO} is not accounted for, resulting in high inter-RIS interference in RIS path separation based on temporal coding, as pointed out in Sec.~\ref{sec_example}. As a result, the quality of the subsequent \acp{AoD} and position estimation significantly deteriorates for non-zero CFO. For the sake of comparison, the \acp{RMSE} achieved by the proposed estimation algorithms and the corresponding \acp{CRB} are also depicted (denoted by the legends 'RMSE-LoS', 'RMSE-NLoS-ML' for the NLoS-ML estimator and 'RMSE-NLoS-LC' for the NLoS-LC estimator),
which shows that using our proposed algorithms, regardless of the value of \ac{CFO}, it is possible to estimate the position with an accuracy matching the theoretical limits, proving the robustness of our proposed algorithms against increasing CFO values. 

\begin{figure}
    \centering
    \begin{tikzpicture}
[scale=1\columnwidth/10cm,font=\footnotesize]
\begin{axis}[
    width=10cm,
    height=7cm,
    ymode=log,
    log ticks with fixed point,
    xmin = 0,
    xmax = 220,
    ymin=0,
    ymax=100,
    xlabel={CFO [Hz]},
    ylabel={RMSE/CRB [m]},  
    grid=major,
    legend style={at={(0,1)},anchor=north west, scale = 1, legend columns = 1, fill opacity=0.5,
            draw opacity=1,   
            text opacity=1},
    legend cell align=left
]
\addplot[draw=red, mark=triangle] [line width=1pt]table[x=nu_CFO, y=pos_LoS, col sep=comma]{tikzfig/data/RMSE_pos_CFO_sens_LoS_nLoS.txt};
\addplot[draw=red, mark=o] [line width=1pt]table[x=nu_CFO, y=pos, col sep=comma]{tikzfig/data/RMSE_pos_vs_CFO_LoS_modified.txt};
\addplot[draw=black, dashed] [line width=1pt]table[x=nu_CFO, y=PEB_LoS, col sep=comma]{tikzfig/data/CRB_pos_CFO_sens_LoS_nLoS.txt};
\addplot[draw=orange, mark=triangle] [line width=1pt]table[x=nu_CFO, y=pos_nLoS, col sep=comma]{tikzfig/data/RMSE_pos_CFO_sens_LoS_nLoS.txt};
\addplot[draw=green!50!black, mark=o] [line width=1pt]table[x=nu_CFO, y=pos, col sep=comma]{tikzfig/data/RMSE_pos_vs_CFO_nLoS1_modified.txt};
\addplot[draw=orange, mark=o] [line width=1pt]table[x=nu_CFO, y=pos, col sep=comma]{tikzfig/data/RMSE_pos_vs_CFO_nLoS2_modified.txt};
\addplot[draw=blue, dashed] [line width=1.5pt]table[x=nu_CFO, y=PEB_nLoS, col sep=comma]{tikzfig/data/CRB_pos_CFO_sens_LoS_nLoS.txt};
\addlegendentry{RMSE-LoS-noCFO}
\addlegendentry{RMSE-LoS}
\addlegendentry{CRB-LoS}
\addlegendentry{RMSE-NLoS-noCFO}
\addlegendentry{RMSE-NLoS-ML}
\addlegendentry{RMSE-NLoS-LC}
\addlegendentry{CRB-NLoS}
\end{axis}
\end{tikzpicture}\vspace{-4mm}
    \caption{\ac{CFO} sensitivity analysis, with and without \ac{LoS}, $P = 35$dBm.}
    \label{fig:CFO_sens}
\end{figure}

\subsection{Sensitivity Analysis to Uncontrolled Multipath} \label{subsec:result_MPC}
Here, we discuss the performance of our estimators in presence of uncontrolled \acp{MPC} in the \ac{LoS} BS-UE channel and the \ac{NLoS} BS-RIS-UE channel.
To account for \acp{MPC}, the BS-UE channel and the channel between the \acp{RIS} and the \ac{UE} are modeled as Rician 
\cite{peng2021ris}. The \ac{BS}-\ac{RIS} path is typically regarded as a LoS path because the \ac{BS} is considered to be directive and the positions of BS and \acp{RIS} are known \cite{abrardo2021intelligent}, \cite{jiang2023two}. Given our scenario with single-antenna \ac{BS} and two \acp{RIS} at different positions, the \ac{BS} cannot be directive towards both \acp{RIS} simultaneously. Thus, a more realistic approach would be to consider \acp{MPC} in the channel between the \ac{BS} and the \acp{RIS} as well. This results in an updated version of the received signal in \eqref{eq_yy}:
\begin{align}\label{eq_yy_MPC}
    &\yy =  \yy_\rmlos^\rmmpc + \sum_{r=1}^2\yy_\rmrisr^\rmmpc+ \nn,
\end{align}
where $\yy_\rmlos^\rmmpc$ denotes the \ac{LoS} path including \acp{MPC}, defined as
\begin{align}
    \yy_\rmlos^\rmmpc = \Psq \alphalos \left( \sqrt{\frac{\kappa_0}{\kappa_0+1}} + \sqrt{\frac{1}{\kappa_0 + 1}} \tilde{h} \right)\bb(\nu)\,,
\end{align}
and $\yy_\rmnlos^\rmmpc$ denotes the aggregation of the RIS paths together with their \acp{MPC}, given by
\begin{align}\label{eq_yy_MPC_RIS}
    &\yy_\rmrisr^\rmmpc =  \Psq \alpha_r \Big( \Big(\sqrt{\frac{\kappa_{Br}}{\kappa_{Br}+1}} \bW_r^\trp  + \sqrt{\frac{1}{\kappa_{Br}+1}}  \bV_r^\trp \Big)\times \nonumber \\ &  \Big(\sqrt{\frac{\kappa_{Rr}}{\kappa_{Rr}+1}}\aab(\thetabr) + \sqrt{\frac{1}{\kappa_{Rr} + 1}} \tilde{\bh}_r \Big)\Big) \odot \bb(\nu).
\end{align}
Here, $\kappa_0$, $\kappa_{Br}$, $\kappa_{Rr}$ are the Rician factors and $\tilde h \sim \mathcal{CN}(0, 1)$, $\tilde\bh_{Br} \sim \mathcal{CN}(\mathbf{0}, \bI)$, $\tilde \bh_r \sim \mathcal{CN}(\mathbf{0}, \bI)$ are the \ac{MPC} (excluding the direct path) in the \ac{BS}-\ac{UE}, \ac{BS}- $\thn{r}$\ac{RIS} and $\thn{r}$\ac{RIS}-\ac{UE} paths respectively, and $\bV_r \triangleq [\tilde\bh_{Br} \odot \gammabrf{0} ~ \cdots ~ \tilde\bh_{Br} \odot \gammabrf{M-1}] \in \complexset{N}{M}$. For sufficiently large $\kappa_0$, $\kappa_{Br}$ and $\kappa_{Rr}$, \eqref{eq_yy_MPC} will converge to \eqref{eq_yy}. The details on how \eqref{eq_yy_MPC_RIS} is derived are explained in Appendix~\ref{app:MPC}.

Fig.~\ref{fig:pos_vs_kappa} shows the effect of \ac{MPC} on our estimators, in which we take $\kappa_{Br} = \kappa_{Rr} = \kappa_{0} = \kappa$ for the sake of simplicity. The transmit power is chosen to be $P = 35$ dBm. It can be observed that with $\kappa$ as small as $10$, sub-dm accuracy can be achieved and that the proposed algorithms can attain near-optimal performance in the sense of converging to the bounds when the direct paths in the BS-RIS, RIS-UE and BS-UE channels are $20$ dB stronger than \acp{MPC} (i.e., when $\kappa = 100$).

Operating at a frequency of 30 GHz, we expect the Rician factor to be significant, indicating that the LoS and RIS paths will dominate and that the MPC will generally be weak. 
However, we must acknowledge that in scenarios where the Rician factor is low, the stochastic nature of NLoS components can lead to performance degradation. For example, strong ground reflections from the RIS may introduce substantial interference, complicating the resolution of multipath effects. This limitation underscores the importance of considering environmental conditions when interpreting results. Thus, while our findings are promising under typical conditions, they also highlight the challenges posed by strong multipath reflections and the inherent limitations of our approach.

\begin{figure}
    \centering

\begin{tikzpicture}
[scale=1\columnwidth/10cm,font=\footnotesize]
\begin{axis}[
    width=10cm,
    height=7cm,
    ymode=log,
    log ticks with fixed point,
    xmode=log,
    log ticks with fixed point,
    xmin=0.1,
    xmax=100,
    ymin=0.007,
    ymax=35,
    xlabel={$\kappa$},
    ylabel={RMSE/CRB [m]},  
    grid=major,
    legend style={at={(1,1)},anchor=north east, scale = 1},
    legend cell align=left
]
\addplot[draw=red, mark=o] [line width=1pt]table[x=kappa, y=pos, col sep=comma]{tikzfig/data/RMSE_LoS_kappa_MPC.txt};
\addplot[draw=black, mark = asterisk] [line width=1pt]table[x=kappa, y=PEB, col sep=comma]{tikzfig/data/CRB_LoS_kappa_MPC.txt};
\addplot[draw=green!50!black, mark=o, dashed, mark options = {solid}] [line width=1pt]table[x=kappa, y=pos, col sep=comma]{tikzfig/data/RMSE_nLoS1_kappa_MPC.txt};
\addplot[draw=orange, mark=o] [line width=1pt]table[x=kappa, y=pos, col sep=comma]{tikzfig/data/RMSE_nLoS2_kappa_MPC.txt};
\addplot[draw=blue, mark = square] [line width=1pt]table[x=kappa, y=PEB, col sep=comma]{tikzfig/data/CRB_nLoS2_kappa_MPC.txt};
\addlegendentry{RMSE-LoS}
\addlegendentry{CRB-LoS}
\addlegendentry{RMSE-NLoS-ML}
\addlegendentry{RMSE-NLoS-LC}
\addlegendentry{CRB-NLoS}

\end{axis}
\end{tikzpicture}\vspace{-8mm}
    \caption{Position estimation performance in presence of MPC vs. $\kappa$ at $P = 35$ dBm.}
    \label{fig:pos_vs_kappa}
\end{figure}

\subsection{Sensitivity Analysis to UE Velocity} \label{subsec_velocity}
Here, we analyze the performance of our estimators in the presence of \ac{UE}
mobility with velocity $\vv$. To account for the impact of UE motion, we extend the received signal model in \eqref{eq_yy} as
\begin{align} \label{eq_yy_velocity}
     \yy = \Psq \Big( \alphalos \, \bb(\nulos_0) + \sum_{i=1}^{R} \alpharisi \, (\bW_i^\trp \aab(\thetabi)) \odot \bb(\nurisi)  \Big) + \nn ~,
\end{align}
where 
$\nulos_0 = \vvbs/\lambda + \cfo , \vvbs = \vv^\trp (\ppbs - \pp_0)/\norm{\ppbs - \pp_0},$ and $  \nurisi = \vvrisi/\lambda + \cfo ~,
\vvrisi = \vv^\trp (\pprisi - \pp_0)/\norm{\pprisi - \pp_0}$ and $\pp_0$ represents the UE's initial position. In Fig. \ref{fig:pos_vs_vel}, we illustrate how the performance of our estimators varies with UE speed $\norm{\vv}$ for $R=2$. The direction of the UE velocity is fixed to $\vv/\norm{\vv} = [1, 0, 0]^\trp$. As expected, our algorithm performs well only for very low UE speeds, as our frugal setup is specifically designed for localizing a stationary UE. Higher UE mobility introduces errors that our approach is not optimized to handle. Hence, the gap between the RMSE and the CRB increases with increasing $\norm{\vv}$ due to mismatch between the assumed model in \eqref{eq_yy} and the true one in \eqref{eq_yy_velocity}. According to \cite{richards2005fundamentals}, the Doppler-induced phase variation over the $M$ transmissions is negligible when the condition $\lvert M \Ts(\pp_i - \pp_0)^\trp \vv/(\lambda\norm{\pp_i - \pp_0})\rvert < 1/8$ holds for all $i \in {\BS, R_1, \dots, R_R}$. Based on the system parameters in Table~\ref{table:sim}, this condition is approximately satisfied for $\norm{\vv} < 1$ m/s, which coincides with the velocity range where our method maintains sub-dm accuracy in Fig.~\ref{fig:pos_vs_vel}.
\begin{figure}
    \centering


\begin{tikzpicture}
[scale=1\columnwidth/10cm,font=\footnotesize]
\begin{axis}[
   width=10cm,
    height=7cm,
    ymode=log,
    log ticks with fixed point,
    xmode=log,
    xmin=0,
    xmax=10,
    ymin=0.01,
    ymax=50,
    xlabel={$\norm{\vv}$ [m/s]},
    ylabel={RMSE/CRB [m]},  
    grid=major,
    legend style={at={(1,1)},anchor=north east, fill opacity=0.5,
            draw opacity=1,   
            text opacity=1, scale = 1},
    legend cell align=left
]
\addplot[draw=red, line width=1pt,mark=o, mark repeat=2] table[x=speed, y=pos, col sep=comma]{tikzfig/data/RMSE_withLoS_vs_vel.txt};
\addplot[draw=black, line width=1pt, mark repeat=2, dashed] table[x=speed, y=PEB, col sep=comma]{tikzfig/data/CRB_withLoS_vs_vel.txt};
\addlegendentry{RMSE}
\addlegendentry{CRB}
\end{axis}
\end{tikzpicture}

\vspace{-5mm}
    \caption{Position estimation performance vs. UE speed at $P = 35$ dBm.}\vspace{-2mm}
    \label{fig:pos_vs_vel}
\end{figure}
 \subsection{Analysis of Ergodic Capacity}
To further assess the performance of the proposed RIS-enabled localization and synchronization system, we analyze the ergodic capacity under practical RIS beamforming based on estimated UE positions. In each trial, the \acp{RIS} are configured to steer their reflections toward an estimated UE position while compensating for the AoA from the \ac{BS} \footnote{In this case, $\gammab_{r,m} =\aab(-\hat{\thetab}_r) \odot \aab(-\phib_r)$ for $m = 0,\cdots,M-1$.}. The SNR is then computed at the true UE position using the resulting beamforming configuration. This process is repeated over 500 independent trials, each corresponding to a different estimated UE position. 
The ergodic capacity is calculated as $\mathbb{E}[\log_2(1+\text{SNR})]$,
with the expectation taken across all trials. Fig. \ref{fig:ergC} presents the ergodic capacity results for the LoS, NLoS-ML, and NLoS-LC scenarios. In the LoS case, the direct path dominates, making localization accuracy less impactful. Under NLoS conditions, however, capacity improves significantly once accurate positioning is achieved. As shown in Fig. \ref{fig:pos_nLoS_vs_PWR}, this occurs at $P = 10$ dBm for the NLoS-ML algorithm and at $P = 32$ dBm dBm for the NLoS-LC algorithm.
\begin{figure}
\centering


\begin{tikzpicture}
[scale=1\columnwidth/10cm,font=\footnotesize]
\begin{axis}[
   width=10cm,
    height=7cm,
    ymode=log,
    log ticks with fixed point,
    xmin=0,
    xmax=40,
    ymin=0.1,
    ymax=200,
    xlabel={$P$ [dBm]},
    ylabel={Ergodic Capacity [bit/sec/Hz]},  
    grid=major,
    legend style={at={(1,1)},anchor=north east, fill opacity=0.5,
            draw opacity=1,   
            text opacity=1, scale = 1},
    legend cell align=left
]
\addplot[draw=red, line width=1pt,mark=o, mark repeat=2] table[x=P, y=Erg_C_LoS, col sep=comma]{tikzfig/data/Erg_C.txt};
\addplot[draw=green!50!black, mark=oplus, mark repeat=2] table[x=P, y=Erg_C_ML, col sep=comma]{tikzfig/data/Erg_C.txt};
\addplot[draw=orange, mark=square, line width=1pt, mark repeat=2] table[x=P, y=Erg_C_LC, col sep=comma]{tikzfig/data/Erg_C.txt};
\addlegendentry{LoS}
\addlegendentry{NLoS-ML}
\addlegendentry{NLoS-LC}
\end{axis}
\end{tikzpicture}

\vspace{-5mm}
    \caption{Ergodic capacity vs. transmit power.}
    \label{fig:ergC}
\end{figure}
\subsection{Analysis of Localization Bound with Increasing Number of \acp{RIS}}
In the previous subsections, we have considered the minimal hardware infrastructure necessary for enabling localization and synchronization. Here, we analyze the impact of increasing the number of \acp{RIS} on the CRB of position estimation, where LoS path exists. The \acp{RIS} are arranged symmetrically on a square centered at the origin, where the \ac{BS} is located. The RIS phase profiles are chosen randomly.
Fig. \ref{fig:PEB_vs_R} shows how the CRB on positioning changes as we add more \acp{RIS}, in the presence of a \ac{LoS} path. The results suggest that adding more \acp{RIS} improves positioning accuracy, especially when going from two to three, where we achieve sub-millimeter precision. This improvement comes from the additional redundant measurements provided by the new \acp{RIS}. However, after reaching six \acp{RIS}, the improvement saturates. This saturation occurs because adding more \acp{RIS} increases the number of unknowns to estimate—the complex channel gains corresponding to the additional \acp{RIS}—and also introduces more propagation paths, which amplifies cross-path interference and further complicates position-domain estimation.
\begin{figure}
    \centering


\begin{tikzpicture}
[scale=1\columnwidth/10cm,font=\footnotesize]
\begin{axis}[
   width=10cm,
    height=7cm,
    ymode=log,
    log ticks with fixed point,
    xmin=2,
    xmax=8,
    ymin=0.0005,
    ymax=0.02,
    xlabel={$R$},
    ylabel={CRB [m]},  
    grid=major,
    legend style={at={(1,1)},anchor=north east, fill opacity=0.5,
            draw opacity=1,   
            text opacity=1, scale = 1},
    legend cell align=left
]
\addplot[draw=red, line width=1pt,mark=o] table[x=R, y=PEB, col sep=comma]{tikzfig/data/pos_ErrorBound_R.txt};
\addlegendentry{CRB}
\end{axis}
\end{tikzpicture}

\vspace{-5mm}
    \caption{Position error bound vs. number of \ac{RIS} at $P = 35$ dBm.}
    \label{fig:PEB_vs_R}
\end{figure}
\begin{remark}
     While our focus is on a frugal single-antenna, single-carrier setup, the proposed algorithms can be extended to systems with multiple antennas and multiple carriers. In such cases, additional measurements such as \ac{AoD} from the LoS path and \ac{ToA} become available, enabling more accurate localization and synchronization. These extensions offer improved accuracy at the cost of increased infrastructure and resource usage.
\end{remark}
\section{Conclusion} \label{sec:conclusion}
In this paper, we proposed a frugal approach for 3-D localization and frequency synchronization of a single-antenna stationary UE using SISO RIS-enabled communication, even without a LoS path. Our method leverages a single-antenna BS and multiple RISs under NB communication, reducing both system complexity and cost.

We developed estimation algorithms and a GLRT-based LoS detector, and showed through simulations that our method attains theoretical performance bounds at moderate to high transmit powers. Accurate CFO estimation proved essential, as neglecting it caused severe degradation. Robustness was confirmed under multipath with Rician factors as small as 10, while mobility analysis showed the approach is best suited for stationary or very low-speed users.

The algorithms offer practical computational complexity, particularly with parallel processing. Overall, our results demonstrate that accurate localization is possible with minimal spectral and hardware resources, supporting sustainable 6G deployments. Future work includes experimental validation and extending the framework to mobile UEs.
\appendices
\section{RIS Phase Profile - Temporal Coding} \label{app:RIS}
 By applying \eqref{eq:YmatReshape} and \eqref{eq:time_coding} we have
\begin{align} \label{yy0_gen_nu}
    &[\yy_{0}]_{k} = \frac{\Psq}{4} \alphalos \big((+1) e^{j2\pi(4k)\Ts\nu} + 
    (+1) e^{j2\pi(4k+1)\Ts\nu} +\nonumber \\ &
    (+1) e^{j2\pi(4k+2)\Ts\nu} +
    (+1) e^{j2\pi(4k+3)\Ts\nu}\big) + \nonumber \\ & \frac{\Psq}{4} \alpha_1  \big(\left(+g_1(\phib_1, \thetab_1, k)\right) e^{j2\pi(4k)\Ts\nu} +  \nonumber \\ &
    \left(-g_1(\phib_1, \thetab_1, k)\right) 
    e^{j2\pi(4k+1)\Ts\nu} +\nonumber \\ &
    \left(+g_1(\phib_1, \thetab_1, k)\right) e^{j2\pi(4k+2)\Ts\nu} + \nonumber \\ &
    \left(-g_1(\phib_1, \thetab_1, k)\right) e^{j2\pi(4k+3)\Ts\nu}\big) + \nonumber \\ &
    \frac{\Psq}{4} \alpha_2 \big(\left(+g_2(\phib_2, \thetab_2, k)\right) e^{j2\pi(4k)\Ts\nu}+  \nonumber \\ &
    \left(+g_2(\phib_2, \thetab_2, k)\right) e^{j2\pi(4k+1)\Ts\nu} +  \nonumber \\ &
    \left(-g_2(\phib_2, \thetab_2, k)\right) e^{j2\pi(4k+2)\Ts\nu} +\nonumber \\ &
    \left(-g_2(\phib_2, \thetab_2, k)\right) e^{j2\pi(4k+3)\Ts\nu}\big) +[\tilde\nn_0]_k,
\end{align}

\begin{align} \label{yy1_gen_nu}
    &[\yy_{1}]_{k} = \frac{\Psq}{4} \alphalos \big((+1) e^{j2\pi(4k)\Ts\nu} + 
    (-1) e^{j2\pi(4k+1)\Ts\nu} +\nonumber \\ &
    (+1) e^{j2\pi(4k+2)\Ts\nu} +
    (-1) e^{j2\pi(4k+3)\Ts\nu}\big) +  \nonumber \\ &\frac{\Psq}{4} \alpha_1 \big(\left(+g_1(\phib_1, \thetab_1, k)\right) e^{j2\pi(4k)\Ts\nu} +   \nonumber \\ &
    \left(+g_1(\phib_1, \thetab_1, k)\right) e^{j2\pi(4k+1)\Ts\nu} +\nonumber \\ & 
    \left(+g_1(\phib_1, \thetab_1, k)\right) e^{j2\pi(4k+2)\Ts\nu} + \nonumber \\ &
    \left(+g_1(\phib_1, \thetab_1, k)\right) e^{j2\pi(4k+3)\Ts\nu}\big) + \nonumber \\ &
    \frac{\Psq}{4} \alpha_2 \big(\left(+g_2(\phib_2, \thetab_2, k)\right) e^{j2\pi(4k)\Ts\nu}+ \nonumber \\ &
    \left(-g_2(\phib_2, \thetab_2, k)\right) e^{j2\pi(4k+1)\Ts\nu} + \nonumber \\ &
    \left(-g_2(\phib_2, \thetab_2, k)\right) e^{j2\pi(4k+2)\Ts\nu} +\nonumber \\ &
    \left(+g_2(\phib_2, \thetab_2, k)\right) e^{j2\pi(4k+3)\Ts\nu}\big) +[\tilde\nn_1]_k,
\end{align}
\begin{align}    \label{yy2_gen_nu}
    &[\yy_{2}]_{k} = \frac{\Psq}{4} \alphalos \big((+1) e^{j2\pi(4k)\Ts\nu} + 
    (+1) e^{j2\pi(4k+1)\Ts\nu} +\nonumber \\ &
    (-1) e^{j2\pi(4k+2)\Ts\nu} +
    (-1) e^{j2\pi(4k+3)\Ts\nu}\big) +  \nonumber \\ &\frac{\Psq}{4} \alpha_1 \big(\left(+g_1(\phib_1, \thetab_1, k)\right) e^{j2\pi(4k)\Ts\nu} +  \nonumber \\ &
    \left(-g_1(\phib_1, \thetab_1, k)\right) e^{j2\pi(4k+1)\Ts\nu} +\nonumber \\ & 
    \left(-g_1(\phib_1, \thetab_1, k)\right) e^{j2\pi(4k+2)\Ts\nu} + \nonumber \\ &
    \left(+g_1(\phib_1, \thetab_1, k)\right) e^{j2\pi(4k+3)\Ts\nu}\big) + \nonumber \\ &\frac{\Psq}{4} \alpha_2 \big(\left(+g_2(\phib_2, \thetab_2, k)\right) e^{j2\pi(4k)\Ts\nu}+ \nonumber \\ &
    \left(+g_2(\phib_2, \thetab_2, k)\right) e^{j2\pi(4k+1)\Ts\nu} + \nonumber \\ &
    \left(+g_2(\phib_2, \thetab_2, k)\right) e^{j2\pi(4k+2)\Ts\nu} +\nonumber \\ &
    \left(+g_2(\phib_2, \thetab_2, k)\right) e^{j2\pi(4k+3)\Ts\nu}\big) +[\tilde\nn_2]_k,
\end{align}
where $[\tilde\nn_r]_k = {1}/{4} \sum_{l=0}^{3}[\bc_r]_l[\nn]_{4k+l}$. 
\section{Fisher Information Matrix Analysis} \label{app:FIM}
Here, we provide more details about the theoretical limits used to assess the estimation algorithms. We use \ac{FIM} analysis and evaluate it for the channel parameters and the positional parameters. The \ac{FIM} for channel parameters can be found as below \cite{kay1993fundamentals}
\begin{align}
    \FF_\text{ch} = \frac{2}{\sigma^2} \sum_{m = 0}^{M-1}\Re\left\{  \frac{\partial [\bz]_m}{\partial \etabch}\left( \frac{\partial [\bz]_m}{\partial \etabch}\right)^{\rm H}\right\},
\end{align}
where $\bz$ is the noiseless part of the received signal in \eqref{eq_yy} and $\etabch \in \{\etabch^\LoS, \etabch^\NLoS\}$. Then, we can translate $\FF_{\rm{ch}}$ to the positional \ac{FIM}, $\FF_{\rm{po}}$, accordingly
\begin{math}
    \FF_{\rm{po}} = \JJ^{\rm T}\FF_{\rm{ch}}\JJ,
\end{math}
where $\JJ$ is the Jacobian matrix with elements 
\begin{math}
    \JJ_{m,n} = {\partial [\etabch]_m}/{\partial[\etab]_n}.
\end{math}
and $\etab\in \{\etab^\LoS, \etab^\NLoS\}$. In case of measurements with \ac{LoS},  $\JJ \in \realset{10}{11}$ and $\FF_{\rm{po}} \in \realsetone{10}$, and in case of measurements without \ac{LoS}, $\JJ \in \realset{8}{9}$ and $\FF_{\rm{po}} \in \realsetone{8}$.

Finally in terms of the error bounds, in case of measurements with \ac{LoS}, the position error bound (PEB) can be calculated by
\begin{align}
    \text{PEB} ={\sqrt{\text{trace}\left([\FF_{\rm{po}}^{-1}]_{7:9,7:9}\right)}},
\end{align}
and the error bound on CFO can be found by
\begin{math}
    \big({\mathbb{E}\left[(\cfo - {\tilde{\cfo}})^2\right]}\big)^{1/2} \geq \big({ [\FF_{\rm{po}}^{-1}]_{10, 10}}\big)^{1/2}.
\end{math}

In case of measurements without \ac{LoS}, PEB can be found using  
\begin{math}   
\text{PEB} =\big({\text{trace}([\FF_{\rm{po}}^{-1}]_{5:7,5:7})}\big)^{1/2},
\end{math}
and the error bound on CFO can be calculated accordingly using\scalebox{1}{
\begin{math}
    \big({ [\FF_{\rm{po}}^{-1}]_{8, 8}}\big)^{1/2}.
\end{math}}
Error bounds on other parameters can be calculated similarly.
\section{Received signal model in the presence of multi-path components} \label{app:MPC}
The received signal at the $\thn{m}$ transmission is
\begin{align}
    y_m = &\Psq \Big( \text{h}_{\text{BS-UE}}+\sum_{r=1}^2 \bh^\trp_{\text{BS-RISr}}(\phibr)\text{diag}(\gammab_{r,m})\bh_{\text{RISr-UE}}(\thetabr) \Big) \nonumber \\ & \times e^{j2 \pi m \Ts \nu} + n_m.
\end{align}
Here, $\text{h}_{\text{BS-UE}}$ is the channel between the \ac{BS} and the \ac{UE}, $\bh_{\text{BS-RISr}}(\phibr)$ is the channel between the \ac{BS} and the $\thn{r}$\ac{RIS} and $\bh_{\text{RISr-UE}}(\thetabr)$ is the channel between the $\thn{r}$\ac{RIS} and the \ac{UE} as below 
\begin{align}
    & \text{h}_{\text{BS-UE}} =  \alphalos \Big(\sqrt{\frac{\kappa_0}{\kappa_0+1}} + \sqrt{\frac{1}{\kappa_0 + 1}} \tilde{h}\Big), \label{eq:htilde}\\
    & \bh_{\text{BS-RISr}}(\phibr) = \alpha_{Br} \Big(  \sqrt{\frac{\kappa_{Br}}{\kappa_{Br}+1}}\aab(\phibr) + \sqrt{\frac{1}{\kappa_{Br} + 1}} \tilde{\bh}_{Br} \Big), \label{eq:hBtilde}\\
    & \bh_{\text{RISr-UE}}(\thetabr) = \alpha_{Rr} \Big(  \sqrt{\frac{\kappa_{Rr}}{\kappa_{Rr}+1}}\aab(\thetabr) + \sqrt{\frac{1}{\kappa_{Rr} + 1}} \tilde{\bh}_r \Big), \label{eq:hRtilde}
\end{align}
and 
\begin{math}
    \alpha_{Br} \alpha_{Rr} = \alpha_r.
\end{math}
Therefore, $y_m$ would be 
\begin{align}
    & y_m = \sqrt{P}\Big( \text{h}_{\text{BS-UE}} + \sum_{r=1}^2 \alpha_{Br} \Big(\sqrt{\frac{\kappa_{Br}}{\kappa_{Br}+1}}[\bW_{r}]_m^\trp \bh_{\text{RISr-UE}}(\thetabr)  \nonumber \\ &   + \sqrt{\frac{1}{\kappa_{Br}+1}}(\tilde{\bh}^\trp_{Br} \odot \gammab^\trp_{r,m}) \bh_{\text{RISr-UE}}(\thetabr)\Big)\Big)e^{j2 \pi m \Ts \nu}+n_m,
\end{align}
where $[\bW_{r}]_m$ is the $\thn{m}$ column of $\bW_r$ as introduced in \eqref{eq_bW}. Therefore, by defining $\bV_{r}$ such that the $\thn{m}$ column is $[\bV_{r}]_m = \tilde{\bh}_{Br} \odot \gammab_{r,m}$, using \eqref{eq:htilde} and \eqref{eq:hRtilde} and concatenating all $M$ measurements, \eqref{eq_yy_MPC_RIS} will be derived.
\balance 
\bibliographystyle{IEEEtran}
\bibliography{IEEEabrv,Sub/ris_doppler}

\end{document}